\documentclass{imsart}
\firstpage{1}
\lastpage{1}

\usepackage{amsthm}
\usepackage{amsmath}
\usepackage{natbib}
\usepackage[colorlinks,citecolor=blue,urlcolor=blue,filecolor=blue,backref=page]{hyperref}
\usepackage{graphicx}

\startlocaldefs

\usepackage{graphicx,booktabs}
\usepackage{comment}

\newcommand{\Norm}{\mathcal{N}}
\newcommand{\betavect}{\boldsymbol{\beta}}
\newcommand{\phivect}{\boldsymbol{\phi}}

\newcommand{\muvect}{\boldsymbol{\mu}}

\newcommand{\wvect}{\mathbf{w}}
\newcommand{\lambdavect}{\boldsymbol{\lambda}}

\def\SPACE{-1mm}

\newcommand{\splitcell}[2][c]{%
 \begin{tabular}[c]{@{}c@{}}\strut#2\strut\end{tabular}%
}

\endlocaldefs

\begin{document}


\begin{frontmatter}
\title{ Assessing Bayesian Nonparametric Log-Linear Models: an application to Disclosure Risk estimation}

\runtitle{Model selection in BNP Log-Linear Models for risk estimation}

\begin{aug}
\author{\fnms{Cinzia} \snm{Carota}\thanksref{addr1}\ead[label=e1]{cinzia.carota@unito.it}},
\author{\fnms{Maurizio} \snm{Filippone}\thanksref{addr2}\ead[label=e2]{maurizio.filippone@eurecom.fr}}
\and
\author{\fnms{Silvia} \snm{Polettini}\thanksref{addr3}\ead[label=e3]{silvia.polettini@uniroma1.it}
}
\runauthor{C. Carota et al.}

\address[addr1]{Dipartimento di Economia e Statistica \\ 
Universit\`a di Torino \\
\printead{e1}\\
}

\address[addr2]{Department of Data Science\\EURECOM, France \\
\printead{e2}\\
}

\address[addr3]{Dipartimento di  Scienze Sociali ed Economiche \\ Universit\`a di Roma ``La Sapienza'' \\
\printead{e3}\\
}

\end{aug}

\begin{abstract}
We present a method for identification of models with good predictive performances in the family of Bayesian log-linear mixed  models with Dirichlet process random effects.  Such a problem arises in many different applications;  here we consider it  in the context of disclosure risk estimation, an increasingly relevant issue raised by the increasing demand for data collected under a pledge of confidentiality. Two different criteria are proposed and  jointly used via a two-stage selection procedure, in a M-open view. The first stage is devoted to identifying a path of search; then,  at the second, a small number of nonparametric models is evaluated through an application-specific score based Bayesian information criterion.  We test our method on a variety of contingency tables  based on microdata samples from the US Census Bureau and the Italian National Security Administration, treated here as  populations,  and carefully discuss its features. This leads us to a journey around different forms and sources of bias along which we show that (i) while based on the so called ``score+search'' paradigm, our method  is by construction well protected from the selection-induced bias,  and (ii) models with good performances are invariably characterized by an extraordinarily simple structure of fixed effects. The complexity of model selection - a very challenging and difficult task in a strictly parametric context with large and sparse tables - is therefore significantly defused by our approach. An attractive collateral result of our analysis are fruitful new ideas about modeling in small area estimation problems, where interest is in total counts over cells with a small number of observations.
\end{abstract}


\end{frontmatter}


\section{Introduction}\label{intro}

Log-linear modeling provides a convenient way of investigating relationships between categorical variables in contingency tables. However,
when the set of classifying variables is large or there are many categories,  the induced  table not only  is  large, but often also sparse, with a huge set of alternative log-linear specifications. This poses challenging theoretical issues in both model fitting   and selection 
 \citep[see e.g.][respectively, and references therein]{fienberg2012,vehtariP}.  In a recent paper \citep{cflp2015}, we proposed a class of Bayesian log-linear models with nonparametric random effects useful to  overcome the above mentioned  problems in disclosure risk estimation, an increasingly relevant issue jointly  raised  by the increasing demand for data  collected under a pledge of confidentiality and refinement of record linkage techniques. 
We  also suggested that likely under this approach model selection may be pursued within a narrower search space. Indeed, here we argue that under this class of models there is room for a novel approach to model selection: we propose a  method aimed at limiting the bias arising in the phase of  model selection, rather than in model fitting as prescribed by the method proposed in  \citet{skinner:shlomo}. 
Our main contribution, presented in Section~\ref{sec:modsel},  is a two-stage model selection procedure  tailored to estimating {\it global} measures of disclosure risk, in a M-open view, i.e. avoiding the unrealistic assumption that the true data generating model is included in the model space. The first stage  is devoted to identify a small number of nonparametric models that are then evaluated through a new measure of predictive performance, essentially an application-specific score based Bayesian information criterion. While based on the so called ``score+search'' paradigm, the proposed selection method 
is by construction well protected from the selection-induced bias, a serious and ubiquitous problem, often neglected in the literature, but recently 
 emphasized especially in machine learning journals \citep[see, e.g.,][]{Reunanen03,VarmaS,CawleyT07, CawleyT10}. See also \citet{vehtariP}. We  discuss this and other forms 
of bias and the distinctive features of our selection method  both in the context of the recent literature on predictive methods for model assessment, selection and comparison \citep{vehtariOj, GelmanHV,  US16, vehtariP} and in relation to the specialized literature on model selection for disclosure risk estimation \citep{forster:webb, skinner:shlomo, mvreiter:jasa12, manriquevallier:reiter:13}. An attractive collateral result  of our analysis 
 are ideas fruitfully applicable in different fields.  Focusing  on the relationship between good global risk estimates and good per-cell risk estimates, 
 we   show  that 
 overfitting nonparametric log-linear models for the purpose of global risk estimation can be advantageously exploited in specific
 applications where the interest is in  total counts over cells with  small sample frequencies, as, for instance, small area estimation. This is  discussed in Section~\ref{sec:tab:piccole1}.    

The remainder of the paper is structured as follows. 
 Section~\ref{intro1} 
introduces the problem of data confidentiality and 
reviews current approaches to model selection for disclosure risk  estimation.
 Section~\ref{sec:npcorrection}  provides  materials and motivations for  the model selection procedure presented in  Section~\ref{sec:modsel} and suggests the application  sketched in  Section~\ref{sec:tab:piccole1}. There, based on a variety of explorations of  different models over different real data tables, first, we show that nonparametric random effects are a powerful adaptive remedy against the positive bias (that is, overestimation of risks) found by \citet{skinner:shlomo} under insufficiently rich, under-fitting log-linear parametrizations;   successively, we illustrate  in terms of per-cell risk estimates the  nature of such corrections for overestimation of  global risks, thereby finding further interesting applications of the class of models under consideration.    

\section{Data confidentiality and model selection for disclosure risk estimation}\label{intro1}
In meeting the request  to always increase the information content and the detail of the statistical outputs,  Statistical Offices must comply with the legal obligation to protect confidentiality of respondents, targeting at maximizing the analytical content of the released data without disclosing confidential information attached to specific individuals or entities.
We consider the release  of social survey microdata; in this case certain publicly available categorical characteristics of the sampled records (e.g. age, gender, education, geography, household type, \dots) can be used as key variables to match sampled records with units in the population and  disclosure risk can be defined on cells of multi-way contingency tables of such variables. Following the previous literature \citep[see, e.g.,][]{bethlehem,chen:kell,skinner:elliot,skinner:holmes}, we define the risk of re-identification in terms of cells containing one or a small number of individuals, that is ``unique'' and  ``small'' cells, of the sample  and population contingency tables, respectively.
The increasing availability of information from external sources and the request to maximize the detail of the released  variables often imply the presence of a large number of key variables, some with many categories, with the consequence that the number of  cells in the associated contingency table is much larger than the sample size and    
risk assessment in actual data releases requires to handle extremely large and sparse tables.
Data are often protected by reducing  the detail (global recoding) or the number (global suppression) of key variables, which lowers the disclosure risk, but also deteriorates the analytical validity of the data. Identifying the proper protection amounts to finding the proper  balance between disclosure risk and data utility. This requires accurate and repeated estimates of disclosure risk prior to any proposed data release, which, in turn, demand for  ready and safe identification of  good models for risk estimation.

In the literature, risk measures are estimated by introducing suitable models on the contingency tables defined by the key variables. 
Let $F_k$ and $f_k$  be the frequencies in the population and sample contingency tables defined by the key variables, and let $K$ be the total number of cells. Two widely accepted measures of  the  global risk of re-identification, or disclosure risks, are the  number of sample uniques which are also population uniques,  
\begin{equation}
\label{eq:tau1}
\tau_1=\sum_{k=1}^K I(f_k=1, F_k=1)=\sum_{k: f_k=1} I(F_k=1| f_k=1):=\sum_{k: f_k=1} \tau_{1k},
\end{equation} 
and  the expected number of correct guesses if each sample unique is matched with an individual randomly chosen from the corresponding population cell \citep[see, e.g.,][]{rinott:shlomo},
\begin{equation}
\label{eq:tau2}
\tau_2=\sum_{k=1}^K I(f_k=1)\frac{1}{F_k}=\sum_{k: f_k=1} \frac{1}{F_k} :=\sum_{k: f_k=1} \tau_{2k}.
\end{equation}
Often (\ref{eq:tau1}) and (\ref{eq:tau2}) are approximated by $\tau_1^*=\sum_{k=1} ^K I(f_k=1) Pr\{F_k=1|f_k=1\}:=\sum_{k: f_k=1} \tau^*_{1k}$ and $\tau_2^*=\sum_k^K I(f_k=1)E(1/F_k|f_k=1):=\sum_{k: f_k=1} \tau^*_{2k}$, i.e. $E(\tau_i|f_1,...,f_K)$, $i=1,2,$ under the assumption of cell independence.

\citet{skinner:holmes,fienb:makov:1998,carlson,elamir:skinner,forster:webb} and \citet{skinner:shlomo}  introduce a log-linear model for the expected cell frequencies, thereby overcoming the assumption of exchangeability of cells 
\citep[see e.g.][]{bethlehem} 
 implying the unrealistic consequence that risk estimates are constant across cells having the same sample frequencies, but different combinations of key variables. Per-cell estimates may be used to highlight high risk combinations or aggregated to produce an overall,  global, risk measure. 
However, as  mentioned in Section~\ref{intro}, in disclosure risk estimation  log-linear models have almost invariably  to deal  with extremely sparse tables  which pose a number of challenges described,  e.g., in \citet{fienberg2007} and \citet{fienberg2012}.  \citet{mvreiter:jasa12} and \cite{manriquevallier:reiter:13} adopt Bayesian latent structure models that does not suffer from the potential shortcomings of  log-linear models.
  \citet{cflp2015}  suggest instead to overcome them by adopting log-linear models with a standard estimable structure for the parametric fixed effects, specifically the independence structure, whose lack of fit is compensated for by nonparametric random effects described by a Dirichlet Process (DP). 

Model specification for risk estimation is a somehow neglected problem. Here we briefly review the current approaches to model selection, specifically devoted to estimate global measures of disclosure risk.  
 \citet{skinner:shlomo}, recognizing the peculiarity of the inferential problem under consideration, suggest a model search algorithm based on a predictive criterion that we describe next.  Instead, \citet{forster:webb}, restrict their attention to the special sub-family of decomposable graphical log-linear models (thereby avoiding problems in model fitting), and  account for model uncertainty by  averaging inferences over that very specific sub-family. \citet{mvreiter:jasa12} and \citet{manriquevallier:reiter:13}  reconsider the problem, though under a different model specification, as will be discussed in the sequel. 

\citet{skinner:shlomo}  model population and sample frequencies by independent Poisson distributions  with rates $\lambda_k$ and $\pi\lambda_k$, respectively, and  describe the parameters $\lambdavect=(\lambda_1, \dots, \lambda_k, \dots, \lambda_K)$ by a log-linear model, 
\begin{equation} \lambda_k=e^{\mu_k},  \qquad \mu_k= {\wvect'_k} \betavect, 
\label{formula2}
\end{equation} 
where $\pi$ is the sampling fraction supposed to be known, $\wvect_k$ is a $q\times 1$ design vector depending on the values of the key variables in cell $k$ and  $\,\,\betavect$ is a $q\times 1$  vector of fixed effects. 

Clearly,  over-parametrized log-linear models tend to overfit the data for the trivial reason that they tend to the saturated model. 
Specifically,  an exceedingly  complex model 
 induces both bias in  estimators of the Poisson parameters $\lambdavect$ and  an inflation of their variances.    \citet{skinner:shlomo} rely on maximum likelihood (ML) estimates of the log-linear model parameters   $\,\,\betavect$ (and therefore of $\lambdavect$) that they plug into the expressions for the global disclosure risks $\hat{\tau}_i^*, i=1,2,$ under the Poisson model. 
Interestingly,  \citet{skinner:shlomo} notice that the behaviour of such risk estimates evolves regularly with the complexity of  the assumed log-linear specification.  In particular, the bias evolves monotonically from underestimation to overestimation of the global  disclosure risks $\tau_i$, $i=1,2$,  when going from the independence model (I), to the all two-way interactions model (II), to the all three-way interactions model (III), and so on. 
Among such models, they select  the least underfitting as the starting model, and propose a stepwise forward model search  aimed at minimizing the (positive) bias of the corresponding risk estimator. The optimal model is the one that achieves the ``best'' compromise between over- and underestimation according to a minimum error criterion, denoted by  $\hat{B}$\footnote{``We argued that $\hat{B}$ may be viewed as an estimator of the bias of $ \hat{\tau}_i^*, i=1,2,$ in the presence of underfitting, when the bias may be expected to be positive. The properties of $\hat{B}$ in the case of overfitting are more difficult to assess.'';  \citet{skinner:shlomo}, p.993}.   
The authors anticipate that most likely a number of ``reasonable models'' may exist, between which the criterion is not able to discriminate,  and also suggest to use the differences between the  estimates for each of these models as a diagnostic to check the sensitivity of the measures to the model specification  \citep[see][p.994]{skinner:shlomo}. The latter, which  in general is very high   \citep[see also][]{mvreiter:jasa12,fienb:makov:1998},  is found to be small across the reasonably good models, implying a form of robustness. 
\citet{mvreiter:jasa12}  stress that, 
when dealing with large tables,   the models to be compared under  the above approach  amount to several hundreds, and exploration of the whole space becomes unfeasible. 
 They avoid such drawback by abandoning the log-linear formulation, relying instead on a Bayesian version of the Grade of Membership (GoM) model.  This is a latent class model characterized by a pre-specified, small, number of classes 
  (extreme profiles), with individuals being allowed to belong to more than one class simultaneously (mixed membership model).
Model complexity is driven by the number of extreme profiles $E$ ($K$ in their notation), 
and it turns out that risk estimates are extremely sensitive to the value of $E$. However,  as $E$ increases, the estimates  exhibit a typical monotonically decreasing pattern which,
past a given threshold for the number of latent classes, tends to become stable.   
This leads \citet{mvreiter:jasa12} to propose the following empirical model selection strategy: 
starting from a given, small, value for $E$,  progressively increase it by a multiple of 3 or 5, depending on the sample size, until there is evidence of stabilization.  The computational cost of such model selection procedure is avoided  in \cite{manriquevallier:reiter:13} by specifying a large number of classes (50) for the truncated latent class model they adopt to allow for structural zeroes. 
Similarly, \citet{Si:Reiter}, following \citet{dunsonxing2009}, use a mixture of independent multinomial distributions, with the mixture distribution being modelled  by a DP prior. The number of classes is not fixed a priori
, yet for computational convenience the authors resort  to a truncated representation of the Dirichlet process, thereby fixing the number of classes to a large value, analogously to \cite{manriquevallier:reiter:13}. 

In this article we focus on the class of Bayesian log-linear mixed models with DP random effects proposed in \citet{cflp2015} and seek an ``optimal'' model within this class.
Under the same assumptions and notation as in (\ref{formula2}), we model the  
 parameters $\lambda_k$ through a log-linear model with mixed effects: \begin{equation} \lambda_k=e^{\mu_k}, \quad \mu_k= \wvect'_k \betavect + \phi_k , \qquad \phi_k|G \stackrel{i.i.d.}\sim   G, \quad G \sim {\cal{D}}(m, G_0),
\label{formula3}
\end{equation} 
where, all other symbols being as in (\ref{formula2}), $\phi_k$ is a random effect accounting for cell specific deviations, whose distribution function, denoted by $G$, is assumed to be  unknown and a priori distributed according to a DP, $\cal{D}$,  with base probability measure $G_0$ and total mass parameter $m$ \citep{ferguson}.
$G_0$ is the mean of the Dirichlet process, 
 while $m$ controls the variance of the process. Suitable parametric priors are also assigned to $m$ and to the fixed effects $\betavect $.

With a slight abuse of terminology, in order to stress the nonparametric nature of random effects, as in  \citet{cflp2015} we refer to this approach as Bayesian nonparametric  (NP) log-linear modelling, though recognizing that it is in fact a semi-parametric approach under which nonparametric, DP distributed, random effects are added to the log-linear formulation (\ref{formula2}) for the Poisson means.   
In describing our models, we will focus on their parametric and  nonparametric components separately; for instance the nonparametric independence model, that we take as a ``default'' model,  is denoted by NP+I,  to emphasize the probabilistic nature of  random effects (NP) and the structure of its parametric component (I), i.e. of fixed effects, respectively.
The parametric counterparts of  models  (\ref{formula3}), that is, models where parametric random effects are added to $\wvect'_k \betavect$ in (\ref{formula2}),  are analysed in \citet{skinner:holmes} and \citet{elamir:skinner}, who observe a practical equivalence, in terms of risk estimation, with log-linear models without random effects.
In the above case we will speak of {\it parametric log-linear models} (or simply {\it parametric models}, labelled P) and {\it parametric risk estimates} to emphasize the nature of the random effects.
\\ In the next section we reconsider the association found by \citet{skinner:shlomo} between underfitting and over-estimation, while the relation  between overfitting and under-estimation will be discussed in Section~\ref{sec:tab:piccole1}. According to the previous authors, such positive/negative bias is due 
to structural zeroes and sampling zeroes, respectively  \citep[for details see][pp.991-992]{skinner:shlomo}. Although our view about the source of over-estimation is different - 
structural zeroes are removed from our analysis since the beginning \citep[see ][p.535]{cflp2015}-, a careful analysis of both such associations conducted from within the enlarged family of log-linear models (\ref{formula3}) provides fruitful ideas about two very challenging issues: model selection, and small area estimation, respectively.

             \section{Underfitting and over-estimation: a Bayesian nonparametric correction}
             \label{sec:npcorrection}
In this Section, we illustrate how the performance of  simple  models of type (\ref{formula2}) is improved by the addition
of Dirichlet process random effects. 
A series of preliminary explorations reveals that such nonparametric log-linear models attain extraordinarily appropriate corrections of the positive bias of risk estimates 
corresponding to the  models of type (\ref{formula2}) from which they are built. 
 Moreover, good nonparametric  models are invariably found in 
very close neighbourhoods of  the nonparametric independence one, NP+I. Here we present  a selection of these explorations. They serve  not only to highlight the ability of  DP random effects  to correct for the overestimation of global risks typical of oversimplified/underfitting log-linear specifications indicated 
  by   \citet{skinner:shlomo}, 
but also to illustrate the nature of such corrections  in terms of per-cell risk estimates.  Moreover, these explorations  introduce and motivate the model selection procedure proposed in the next Section.  
There we will also provide  a theoretical justification for the model performances  observed in this Section. 
 Models are tested on a range of contingency  tables differing in size, reference population and spanning variables obtained 
from different sources as detailed in the supplementary material A.1. First, we consider three tables of decreasing dimension (K=3,600,000; 900,000 and 360,000 cells, respectively) built from the 5\% public use microdata sample of the U.S. 2000 census for the state of California \citep[IPUMS][]{IPUMS}. To allow for structural zeroes,  we also consider a table of 844,800 cells, half of which structurally empty, built from the set of $N=794,986 $ individuals recorded in the 7\% public use microdata sample of the Italian National Social Security Administration, 2004 (source: Work Histories Italian Panel, WHIP). Such microdata samples are treated here as populations: we take simple random samples from those populations and benchmark estimated risks under different models  against their ``true" values.\\
For each of the four tables above, in this Section we explore a set of nonparametric models 
built as follows. 
We first specify a set of models of type (\ref{formula2})  selected by maximizing the log-likelihood under a severe penalty for complexity.
Of course, these models are not optimal  for disclosure risk estimation, but, conditionally on the severe  constraint of simplicity imposed through the penalty term, they are optimal for estimation of the cell parameters $\lambdavect$.  To each of such models we associate a nonparametric model of type  (\ref{formula3}) by adding DP random effects. A formal description of  the criterion used at this preliminary stage is provided in Section~\ref{sec:modsel}, where it is denoted by $C_0(\gamma)$. 
For each of our test tables, Table~\ref{tab:COMPLEX} (first three rows)  lists the nonparametric models selected for exploration. Models are reported in order of   {\it incremental} complexity w.r.t. our default model, NP+I; complexity is described through:  (a) the number of two-way interactions  (interaction terms)  included  in the model, and  (b) the number of associated interaction parameters  (further details can be found in the supplementary material A.1). As indicated by the entries of column (a) in Table~\ref{tab:COMPLEX},  the models selected through $C_0(\gamma)$  are close neighborhoods of the nonparametric independence one. 
 To illustrate the effect of a richer fixed effects specification,   for the Whip table we also include in the exploration two  models encompassing four two-way interaction terms, corresponding to a much lower penalty for the log-likelihood. All parameters of the models considered in this section are assigned vague priors, as  detailed  in the supplementary material A.1. For  computational details regarding the applications presented here and throghout the article see instead the supplementary material B. 

  \begin{table}[!ht]
\caption{List of nonparametric models explored in California and Whip tables, and their complexity. Complexity is evaluated focusing on the parametric component  and measured through: (a) the number of two-way interactions included  in the model, and  (b) the number of associated interaction parameters. Since our reference model is the nonparametric independence one (NP+I), we restrict attention to the  terms that are added to the main effects of the spanning variables in the fixed effects specification.}
\label{tab:COMPLEX}
\centering
\begin{tabular}{ccccccccc}
\hline
&\multicolumn{6}{c}{California }&\multicolumn{2}{c}{Whip }\\
\cmidrule(lr){2-7}\cmidrule(lr){8-9}
&\multicolumn{2}{c}{{\em Large}: K=3600000.} &\multicolumn{2}{c}{{\em Medium}: K=900000} &\multicolumn{2}{c}{{\em Small}: K=360000}  &\multicolumn{2}{c}{}\\
Label & (a) &(b) & (a) &(b)& (a) &(b)& (a) &(b)\\
\cmidrule(lr){1-1}\cmidrule(lr){2-3}\cmidrule(lr){4-5}\cmidrule(lr){6-7}\cmidrule(lr){8-9}
NP$_a$ &1 &1 & 1& 3  & 1   & 1 & 1     &  9\\
NP$_b$ & 1& 18 &1  & 18 &1 &  20   & 1 & 12\\ 
NP$_c$ &2& 19 & 2  & 21 && &    2   &21\\ 
NP$_d$&&&&&&  & 4   & 27 \\ 
NP$_e$ &&&&&&& 4   &   48\\   
\hline
\end{tabular}
\end{table}
  \begin{figure}[p]
  \centering
 \includegraphics[height=0.4\textwidth, angle=-90]{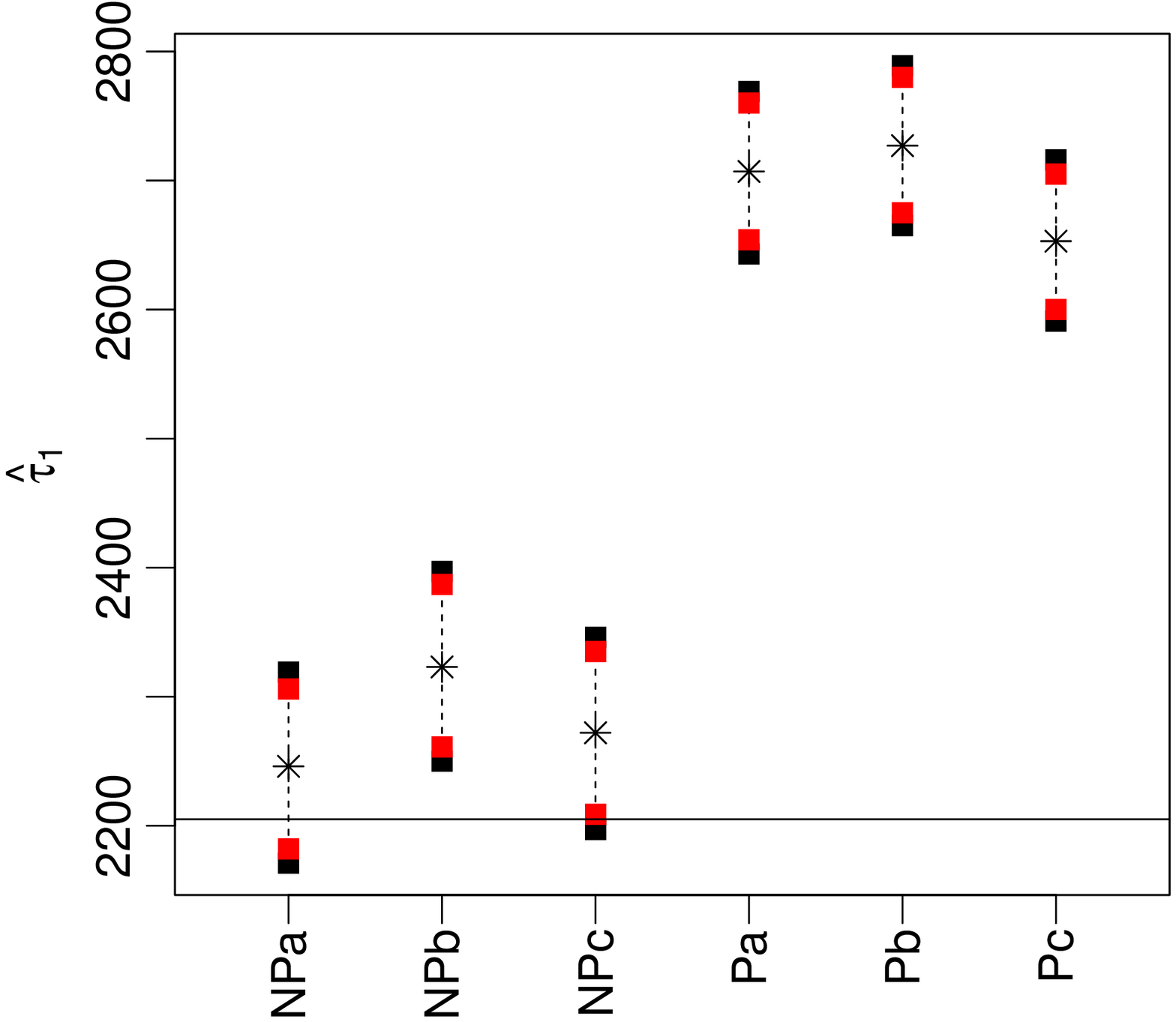}
 \includegraphics[height=0.4\textwidth, angle=-90]{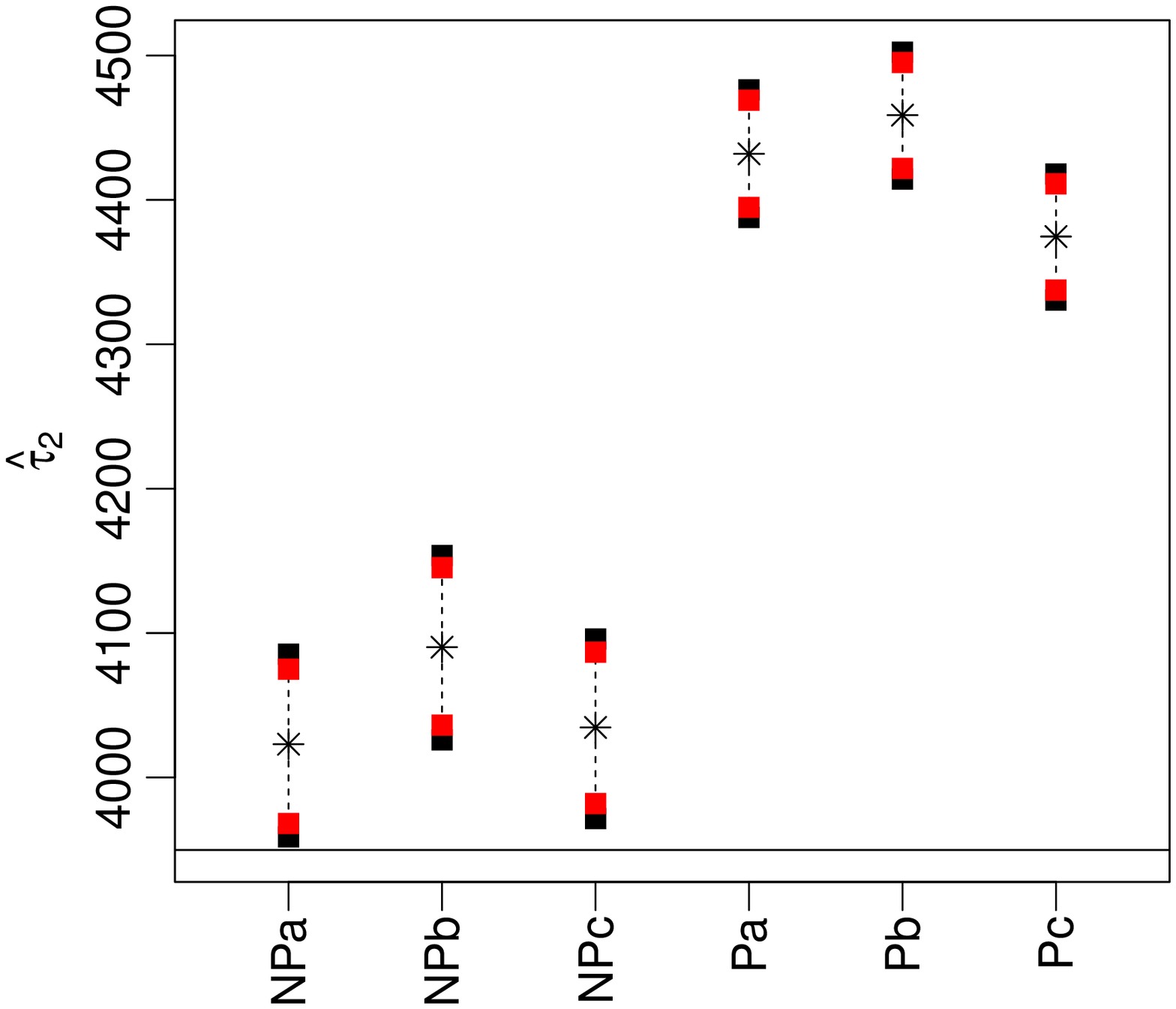}
 \includegraphics[height=0.4\textwidth, angle=-90]{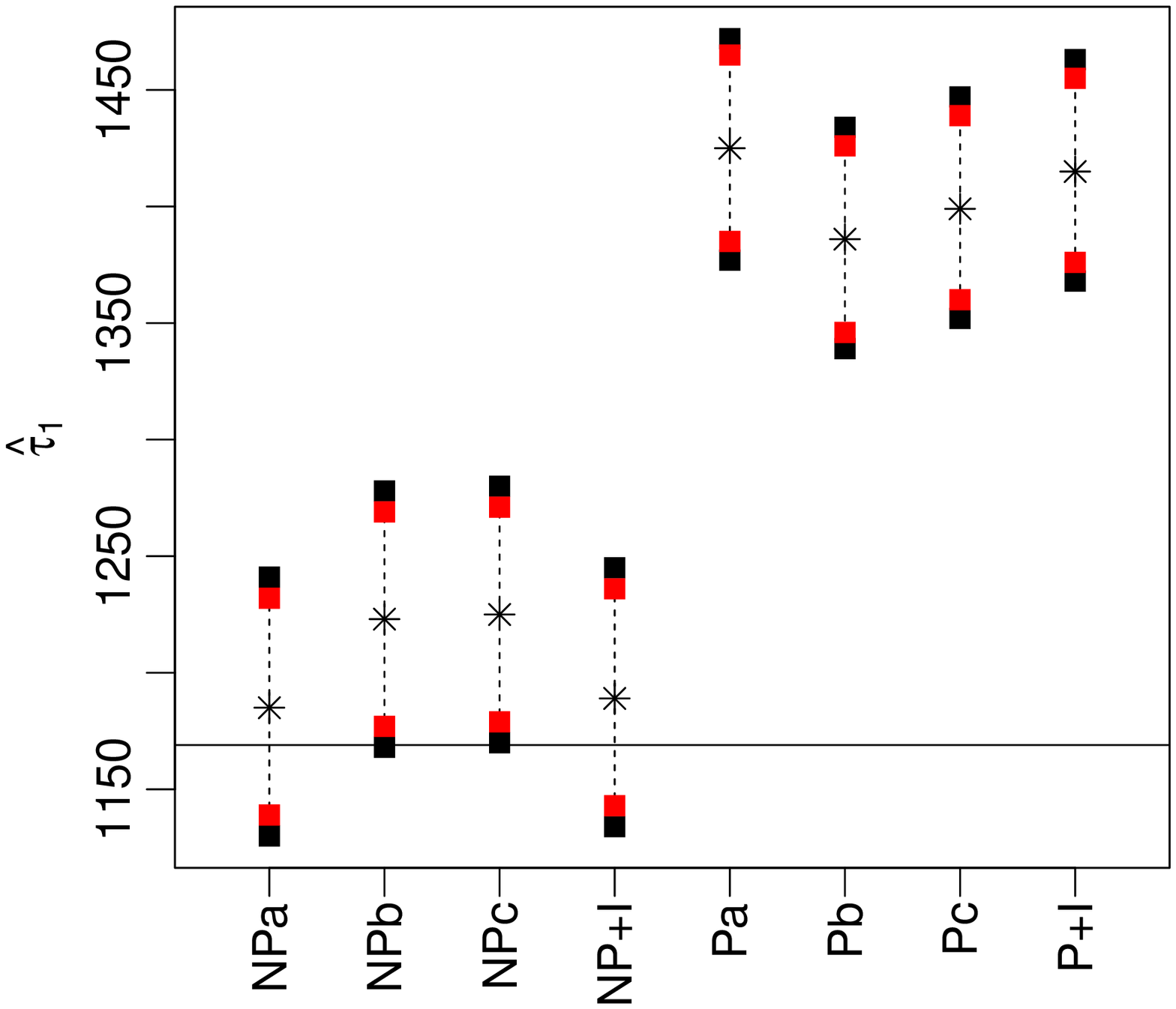}
 \includegraphics[height=0.4\textwidth, angle=-90]{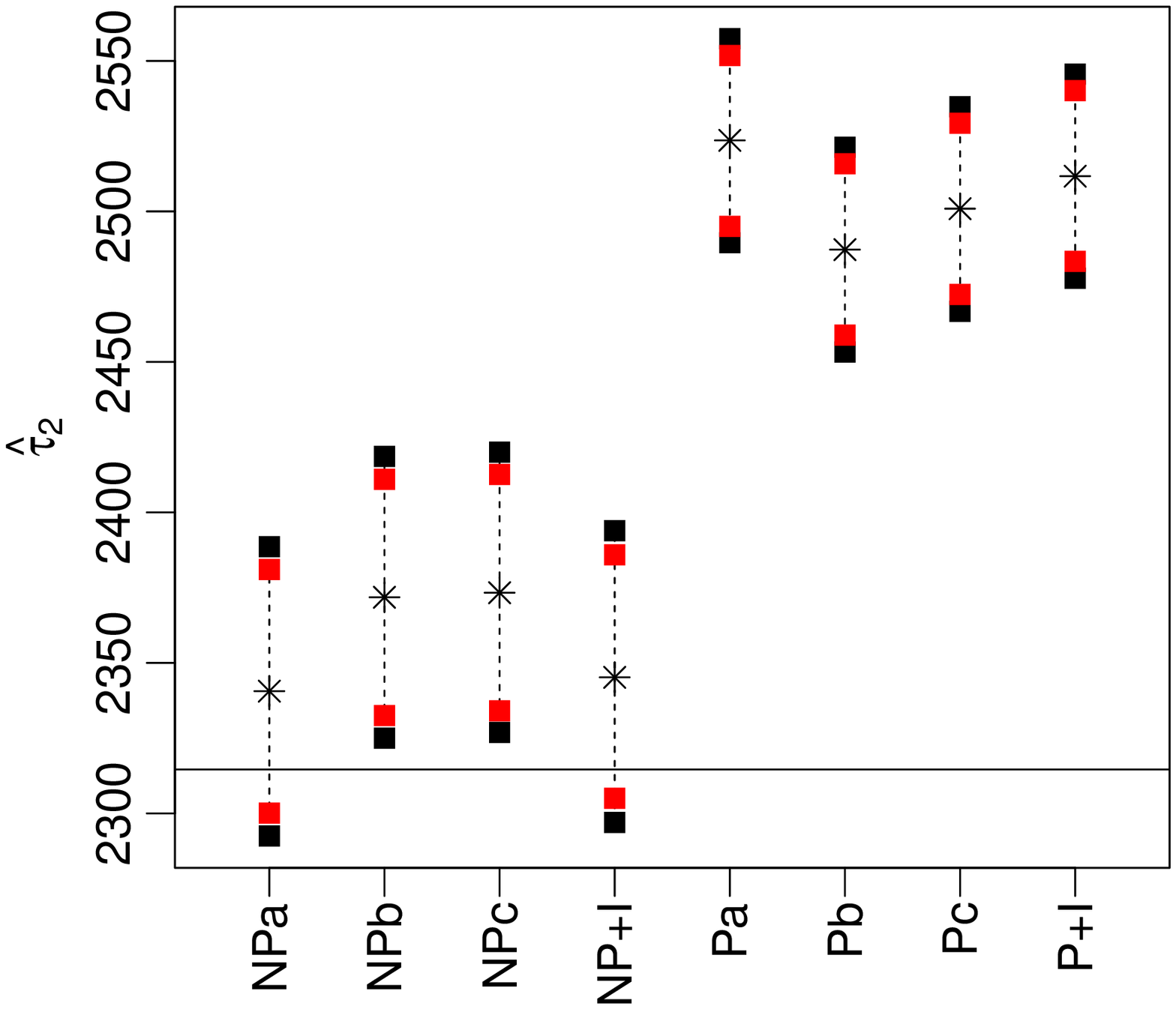}
 \includegraphics[height=0.4\textwidth, angle=-90]{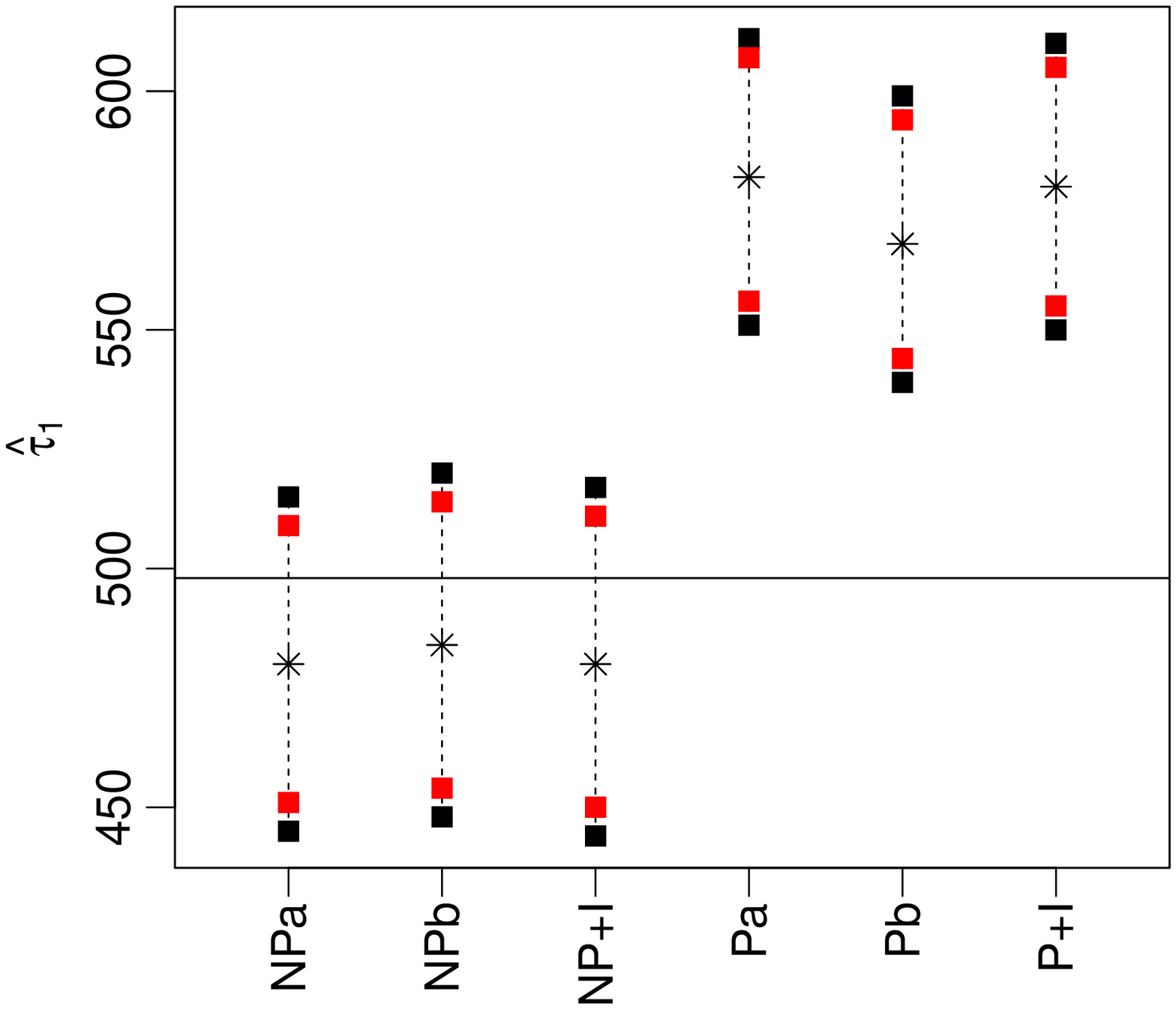}
 \includegraphics[height=0.4\textwidth, angle=-90]{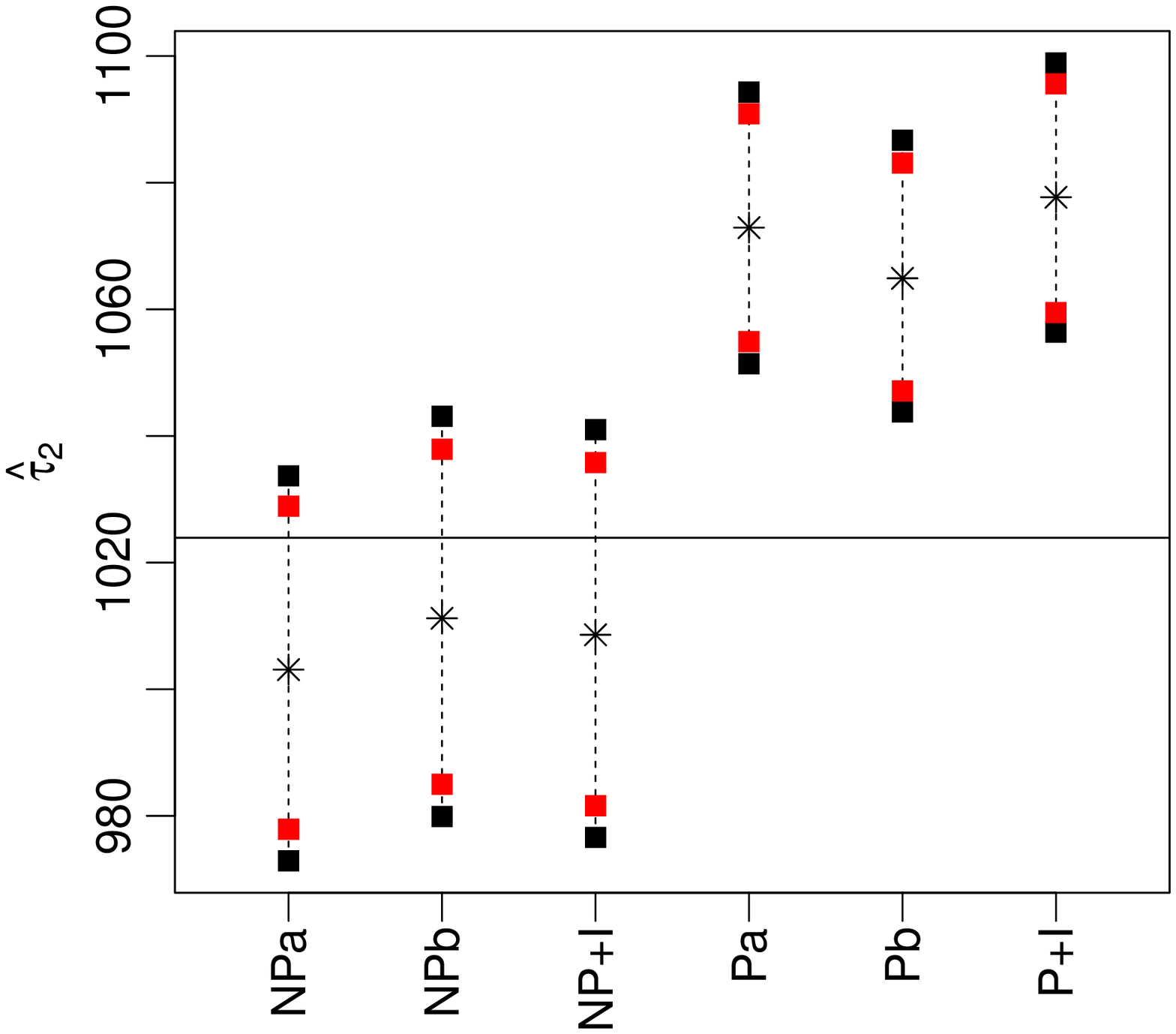}
  \caption{California data: true values (horizontal solid line) and quantiles (0.005, 0.025, 0.50, 0.975, 0.995) of the posterior distributions of ${\tau}_{1}$ (first column) and ${\tau}_{2}$ (second column) under the set of nonparametric  models (NP) reported in Table~\ref{tab:COMPLEX}, and under their  parametric counterparts (P).  First row: {\em Large} table; second row:  {\em Medium} table; third row:  {\em Small} table.  For the latter two tables nonparametric and parametric independence models (NP+I and P+I, respectively) are also reported.}
  \label{fig:quantili:tab1e2}
  \end{figure}
  
\begin{figure}[t]
 \centering
 \includegraphics[height=0.45\textwidth, angle=-90]{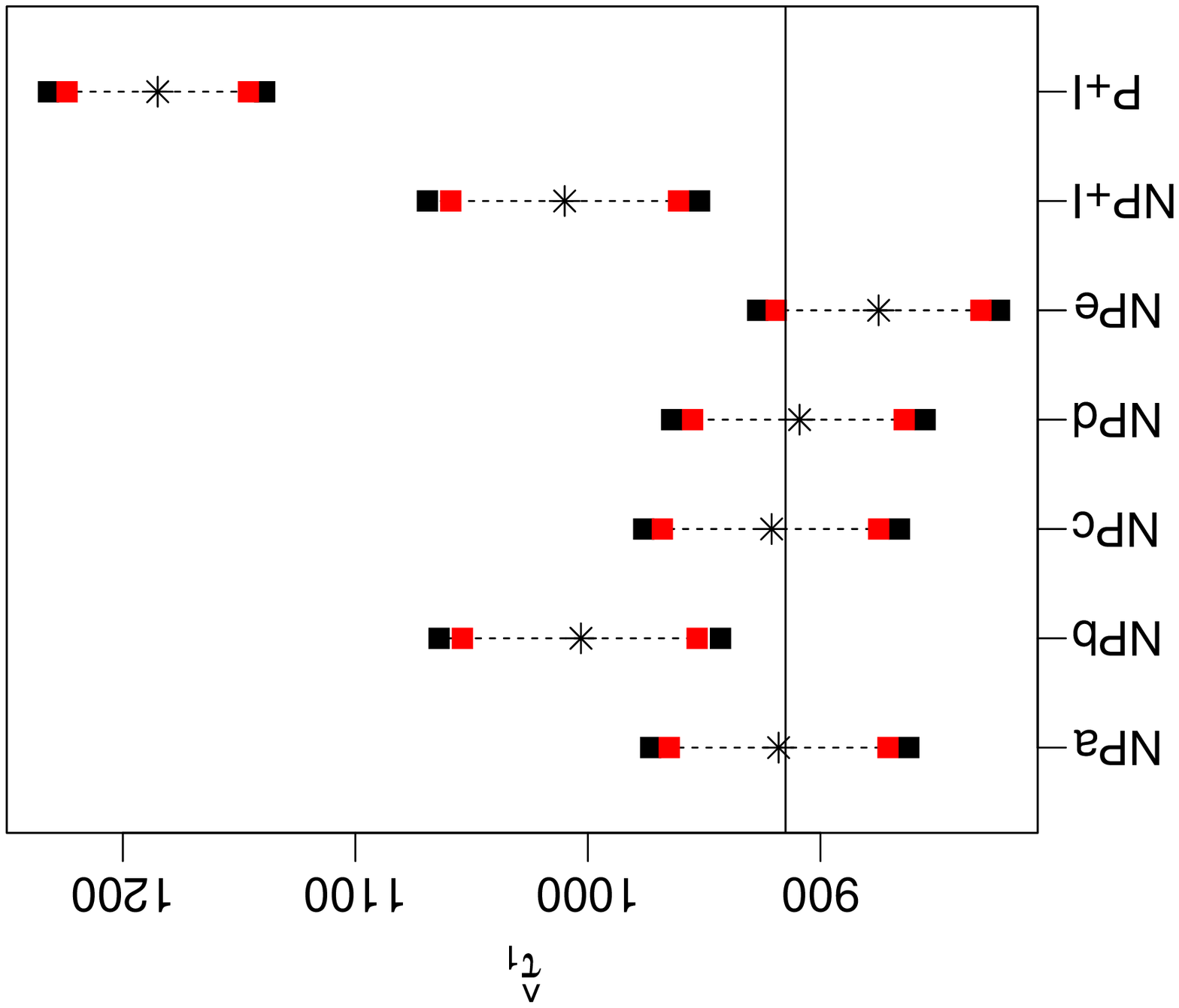}
 \includegraphics[height=0.45\textwidth, angle=-90]{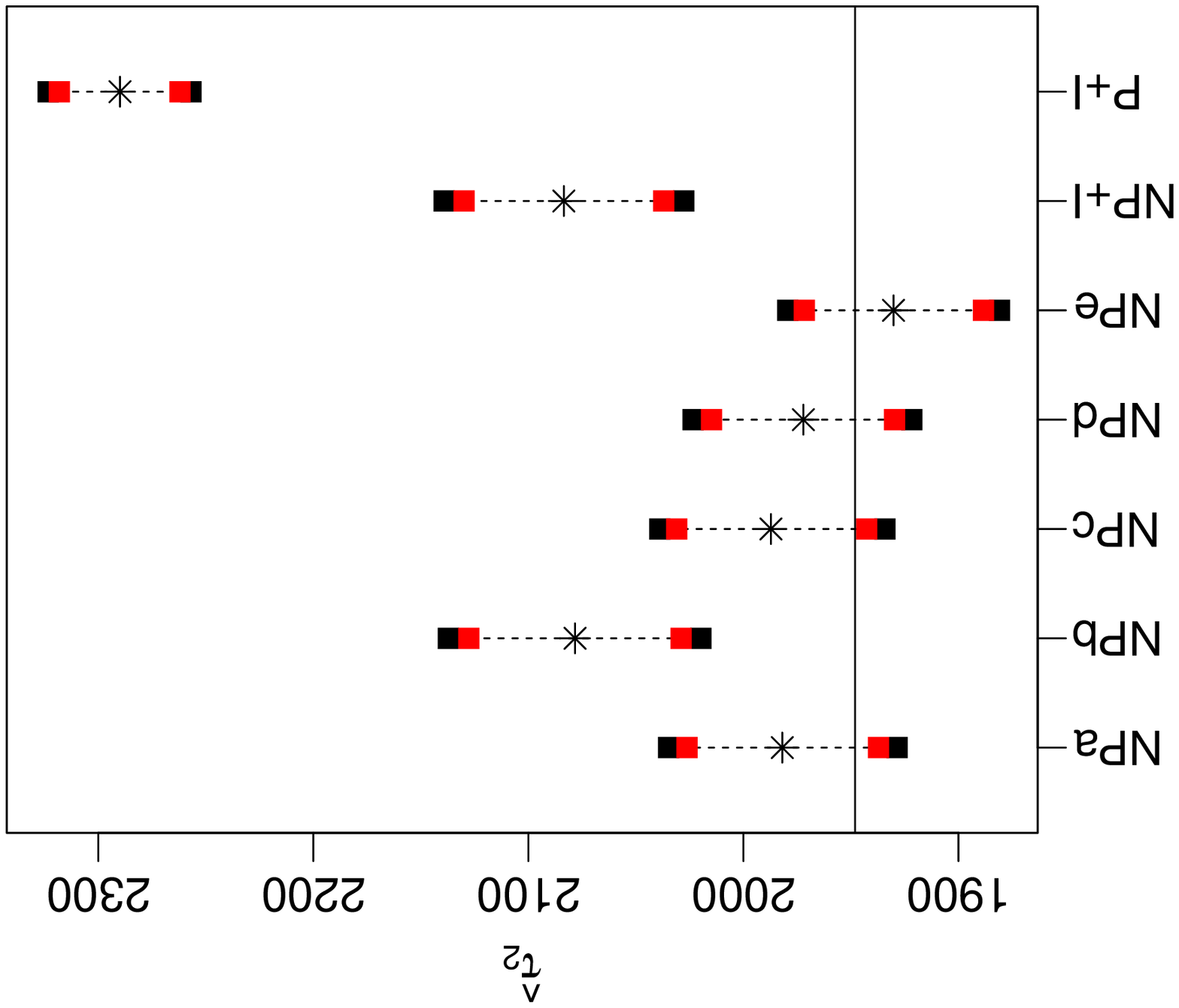}
 \caption{WHIP data:  true values (horizontal solid line) and  quantiles (0.005, 0.025, 0.50, 0.975, 0.995) of the posterior distributions of ${\tau}_{1}$ (left) and ${\tau}_{2}$ (right)  under the nonparametric  models (NP) reported in Table~\ref{tab:COMPLEX} presented in increasing order of complexity. The performance of models is very sensitive to the inclusion of specific interaction terms, rather than to the inclusion of a large number of interaction terms and/or a large number of parameters.  Nonparametric and parametric independence models are also reported. }
 \label{fig:whipmidtau1quantiles}
 \end{figure}

Figures~\ref{fig:quantili:tab1e2} and ~\ref{fig:whipmidtau1quantiles} present posterior medians  and posterior credible intervals  (95\% and 99\%) for the global risks $\tau_1$ and $\tau_2$  under all nonparametric  models presented  in  Table~\ref{tab:COMPLEX}, for California and WHIP tables, respectively.   Figure~\ref{fig:quantili:tab1e2} includes the parametric counterparts  of the above models to assess their relative performance in terms of risk estimates. Clearly, in the presence of DP random effects, good risk estimates stem from a few, simple fixed effects, which are otherwise insufficient to produce adequate inferences in the presence of parametric random effects, or, equivalently \citep{elamir:skinner}, in the absence of random effects under a formulation of type  (\ref{formula2}). Figure~\ref{fig:whipmidtau1quantiles} confirms that  simple nonparametric models are able to produce good risk estimates and suggests the result of selecting a richer fixed effects specification.  Actually, in all of our preliminary explorations we  observed  that the mere inclusion of one specific interaction term may mark the difference between good models and inadequate models. Indeed, the good performance of the four models  $NP_e$,  $NP_d$,  $NP_c$ and  $NP_a$  
 is due to the presence of  a specific interaction term, as we will see in Section~\ref{sec:modsel}.  
Figure~\ref{fig:whipmidtau1quantiles} also presents the risk estimates  obtained under the parametric independence model (P+I), as its performance will be further analysed in the sequel (see  Figure~\ref{fig:sigmoidinuovi:tab1} in the supplementary material A.2).

\begin{figure}[ht]
  \centering
\def\arraystretch{-1}
\begin{tabular}{ccc}
 (a)&(b)&(c)\\
 \includegraphics[height=0.3\textwidth, angle=-90]{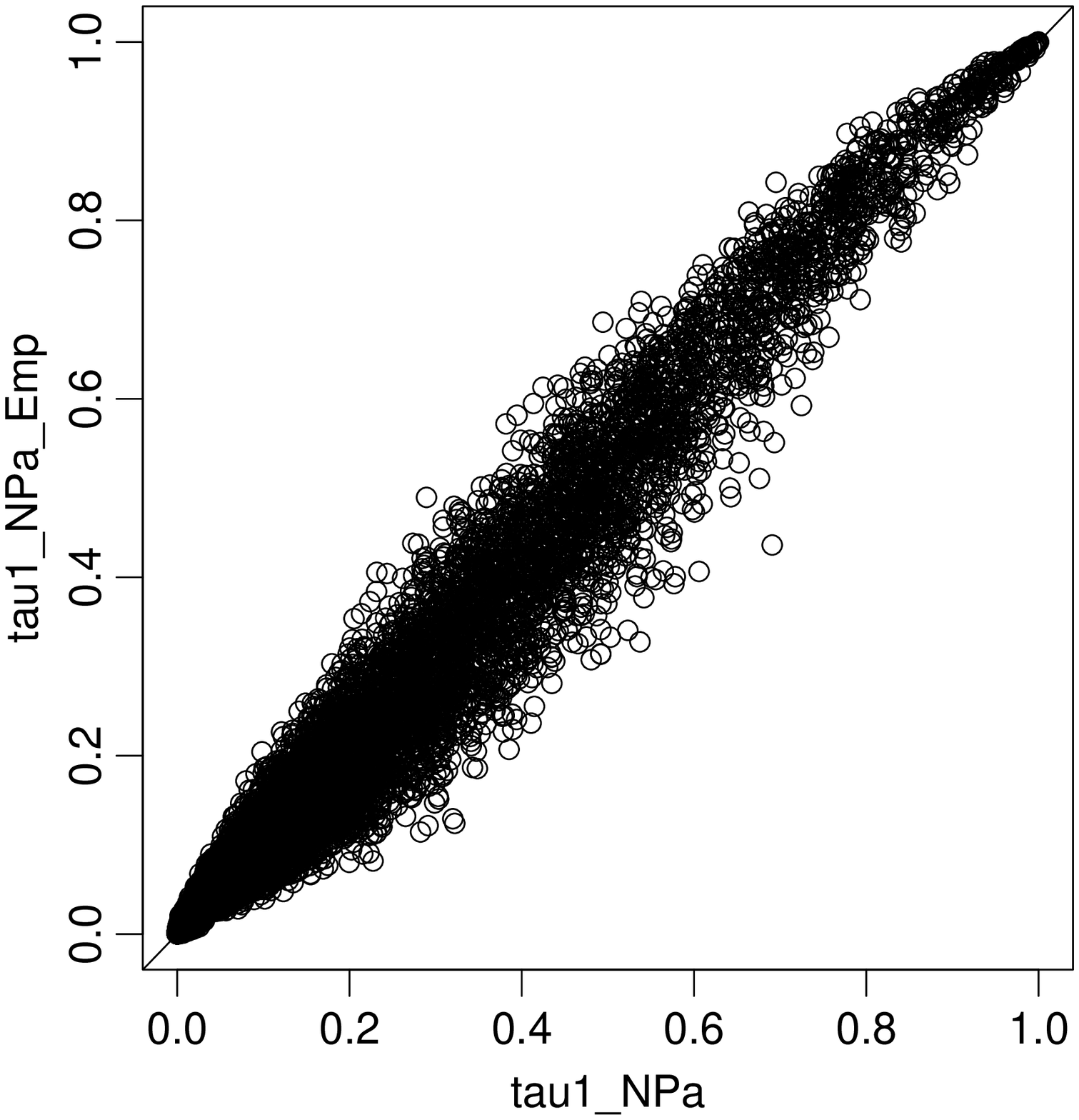} &
 \includegraphics[height=0.3\textwidth,angle=-90]{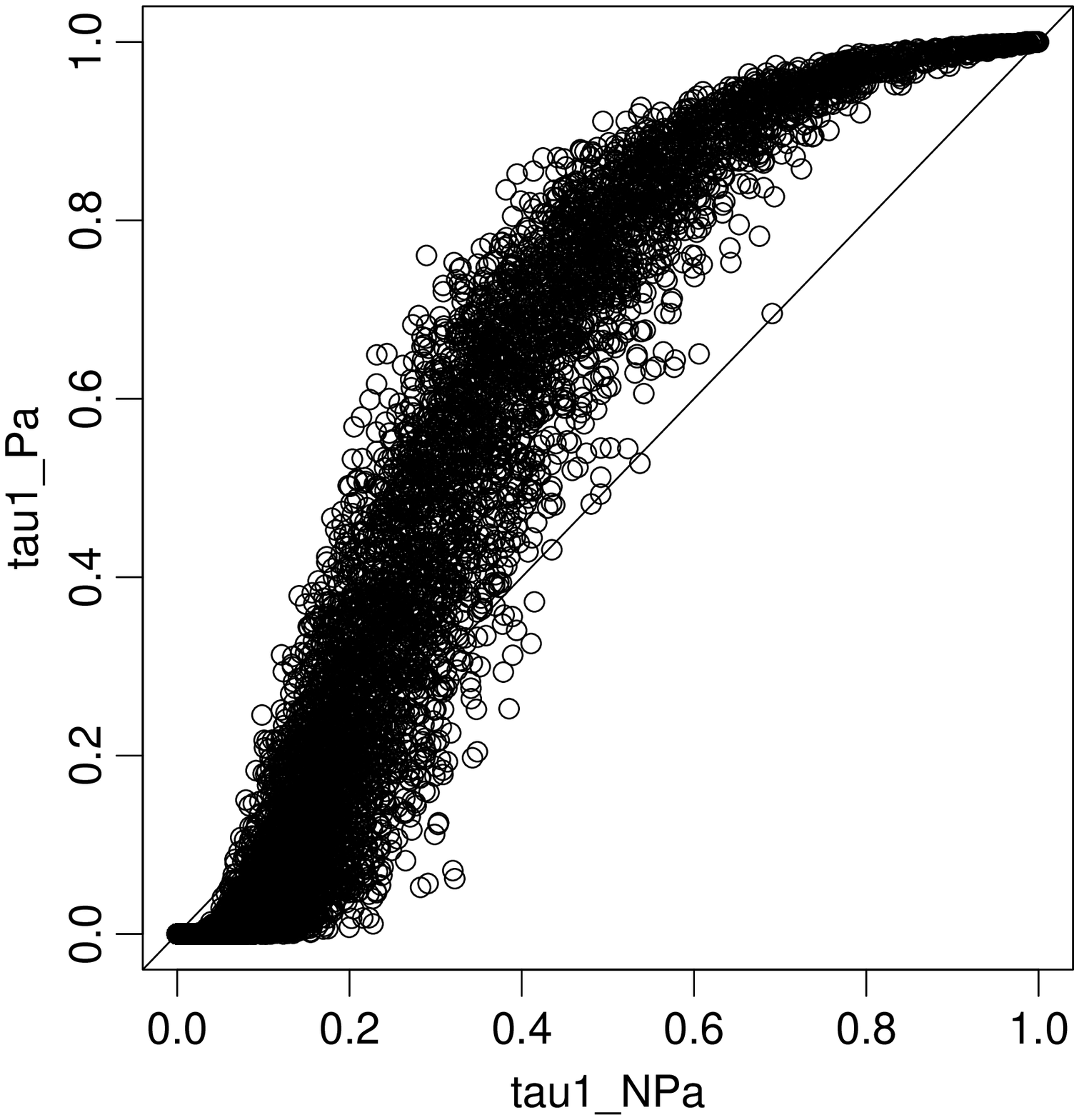} &
 \includegraphics[height=0.3\textwidth,angle=-90]{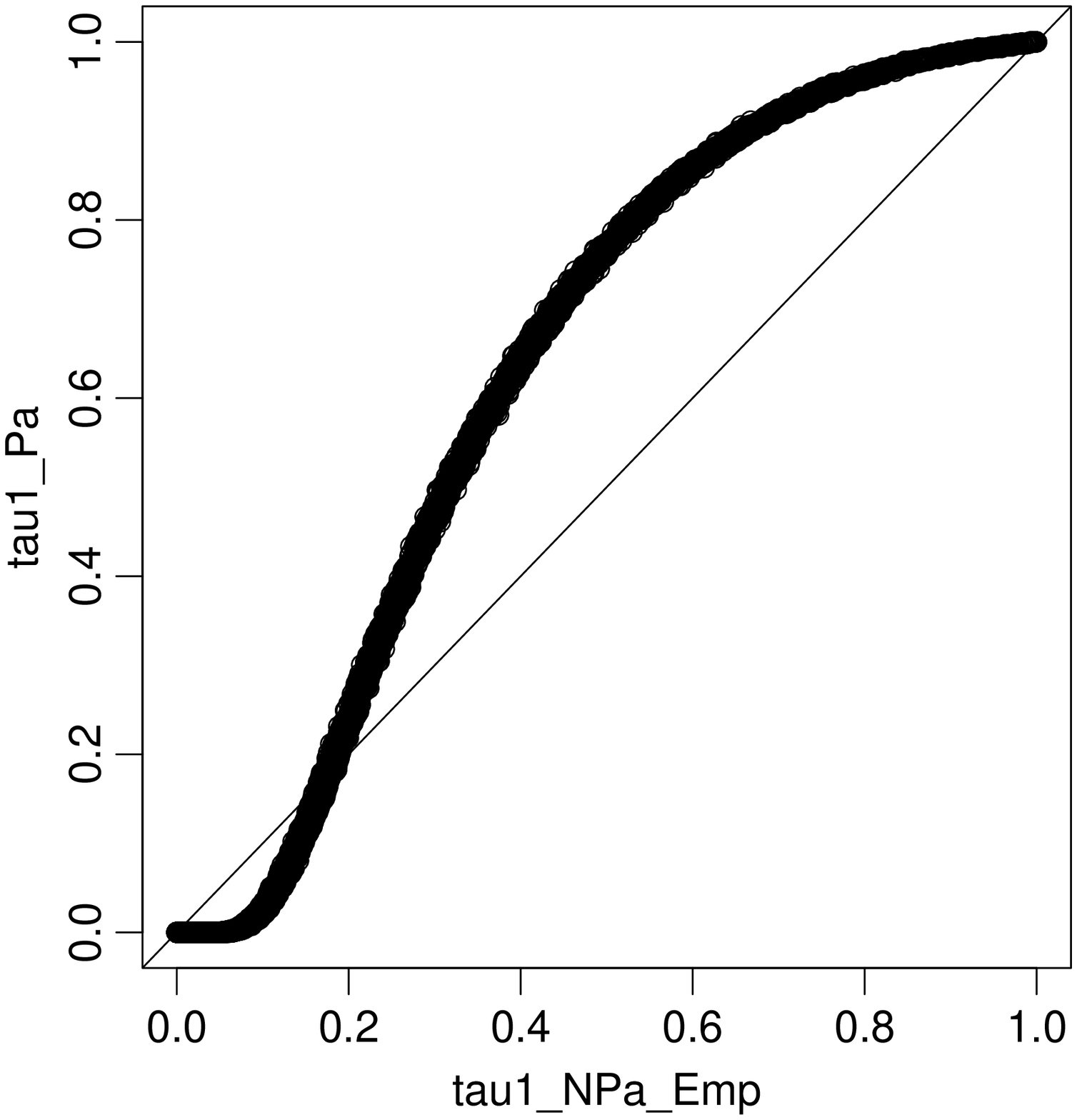}  \\
   \includegraphics[height=0.3\textwidth, angle=-90]{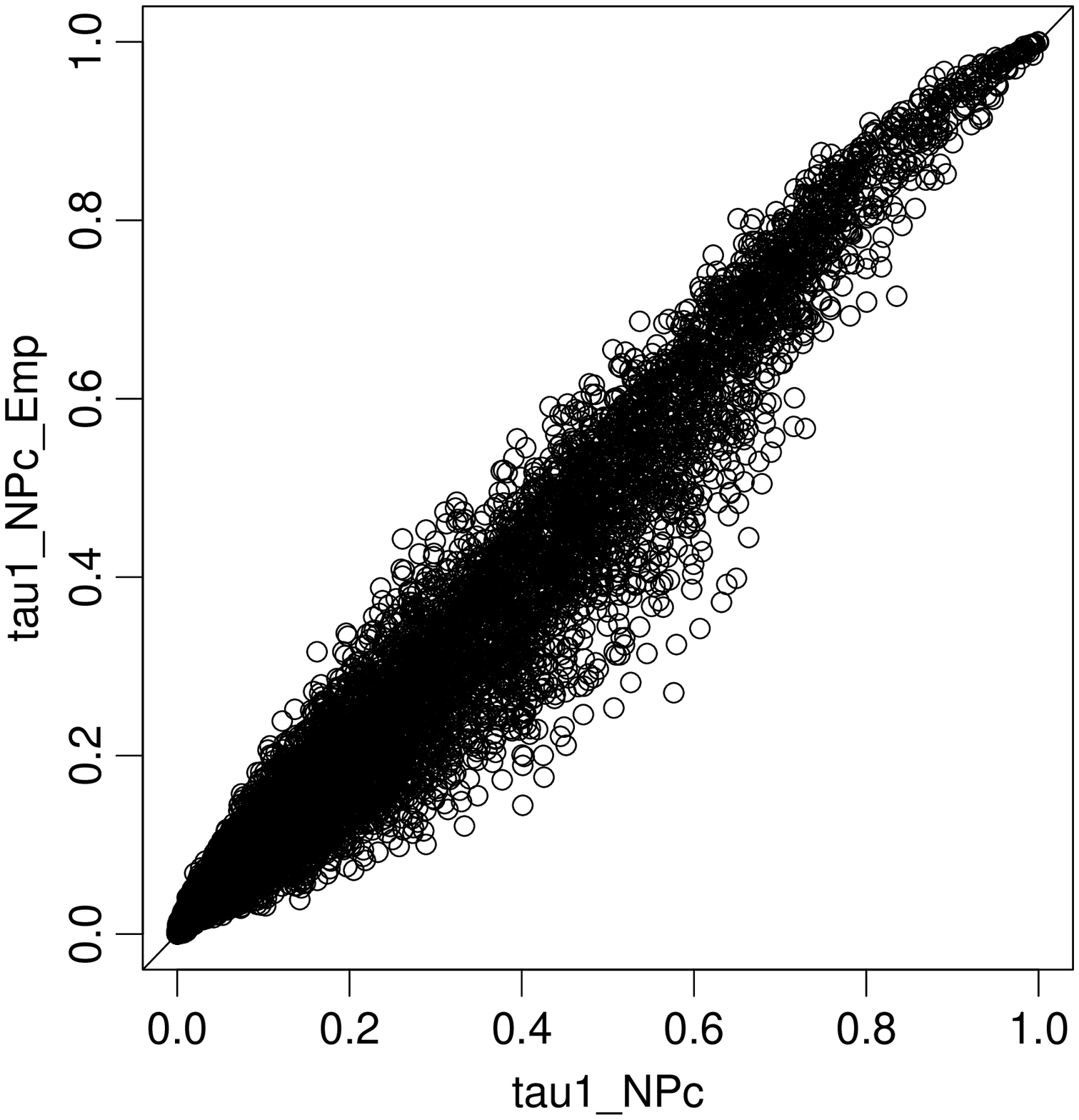} &
     \includegraphics[height=0.3\textwidth,angle=-90]{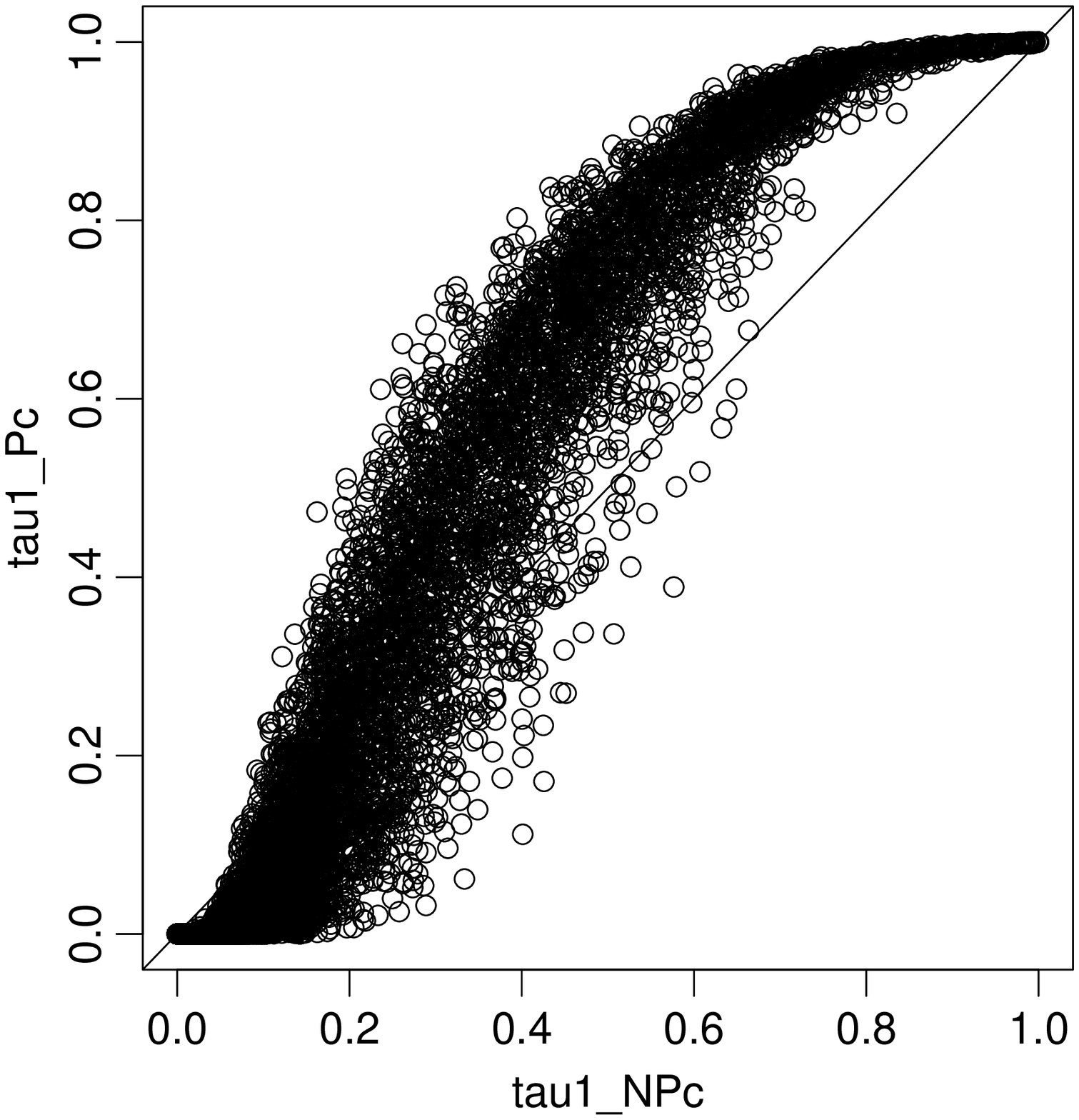} &
     \includegraphics[height=0.3\textwidth,angle=-90]{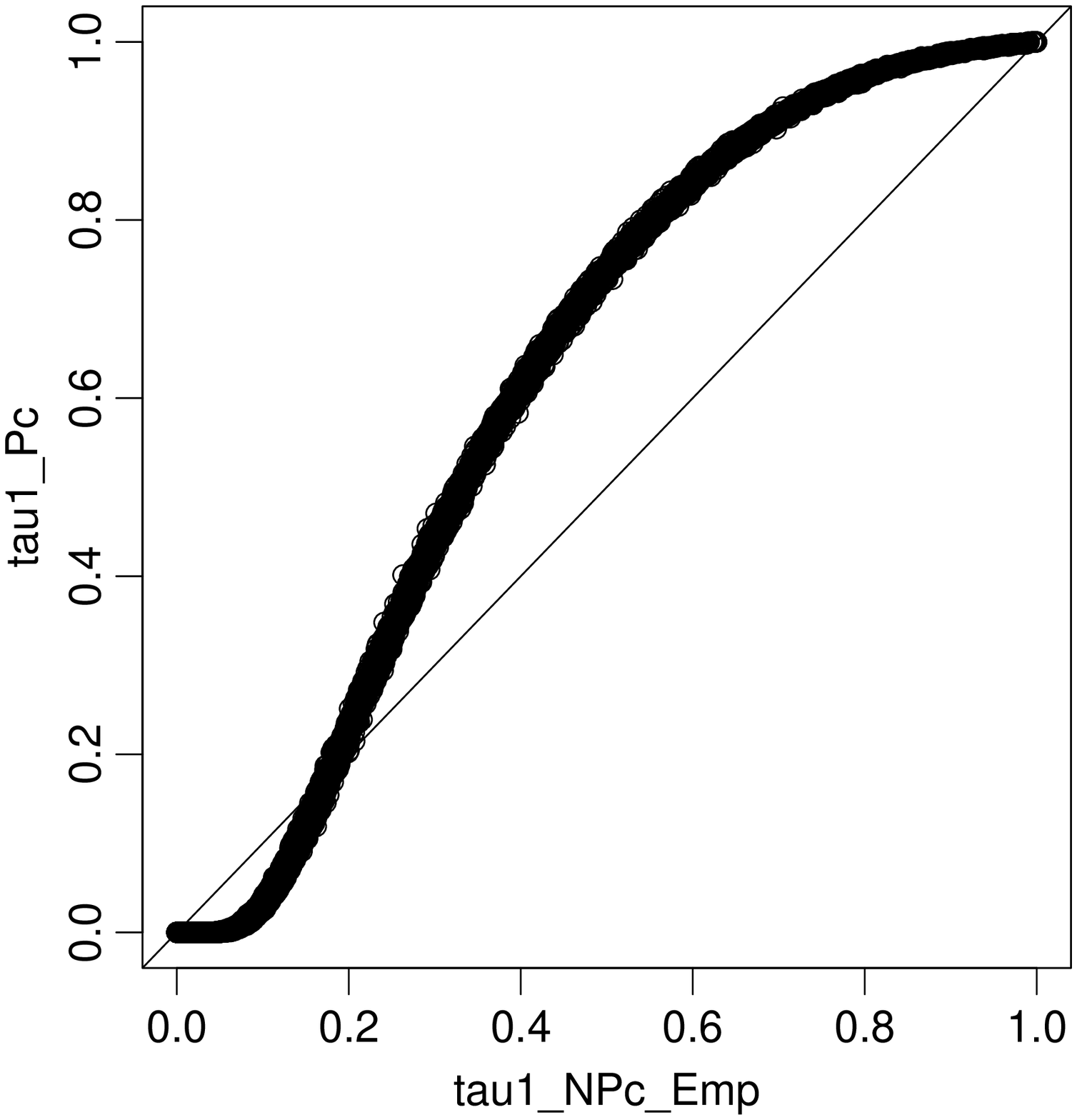}  \\
  \includegraphics[height=0.3\textwidth, angle=-90]{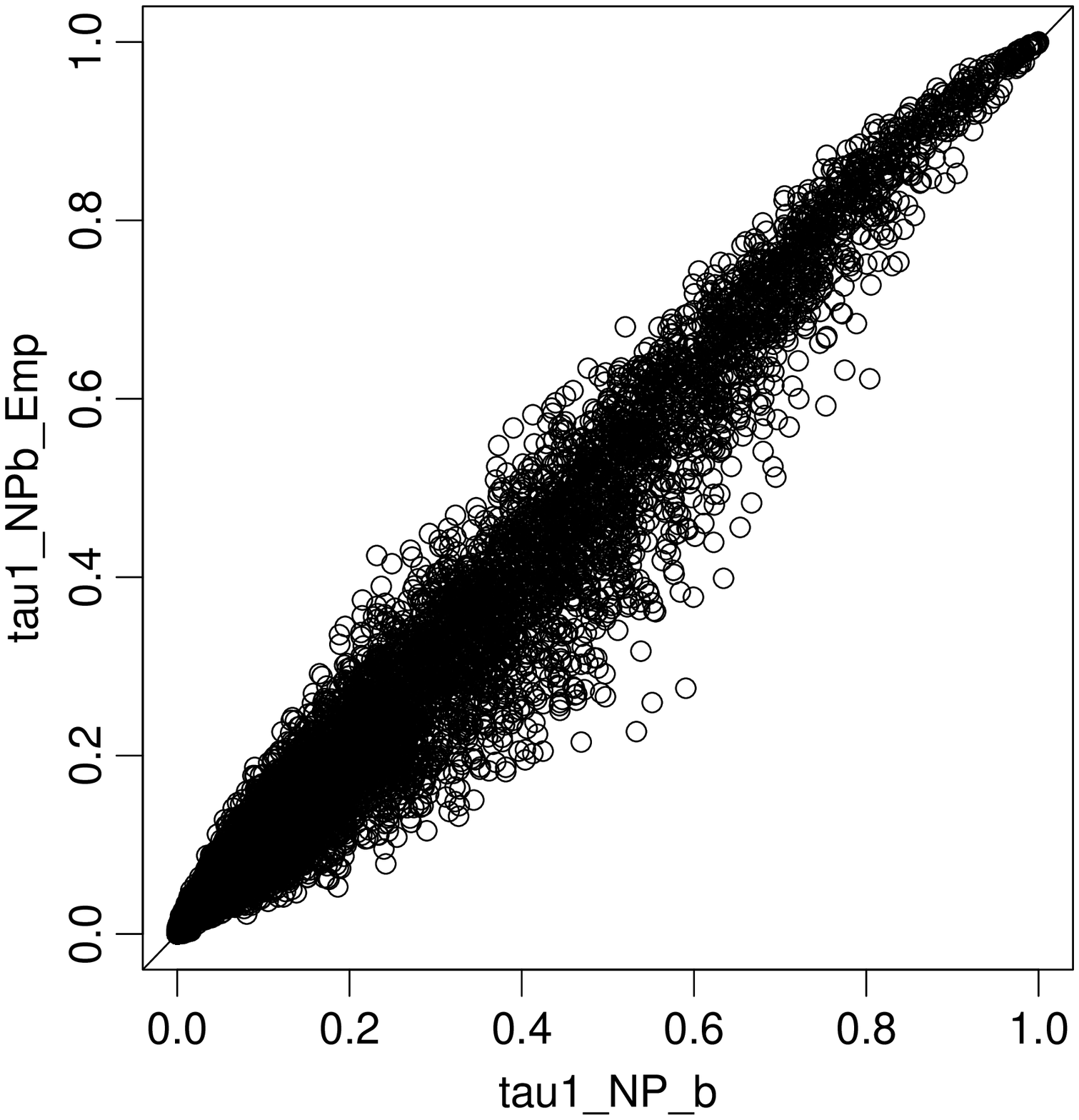} &
   \includegraphics[height=0.3\textwidth,angle=-90]{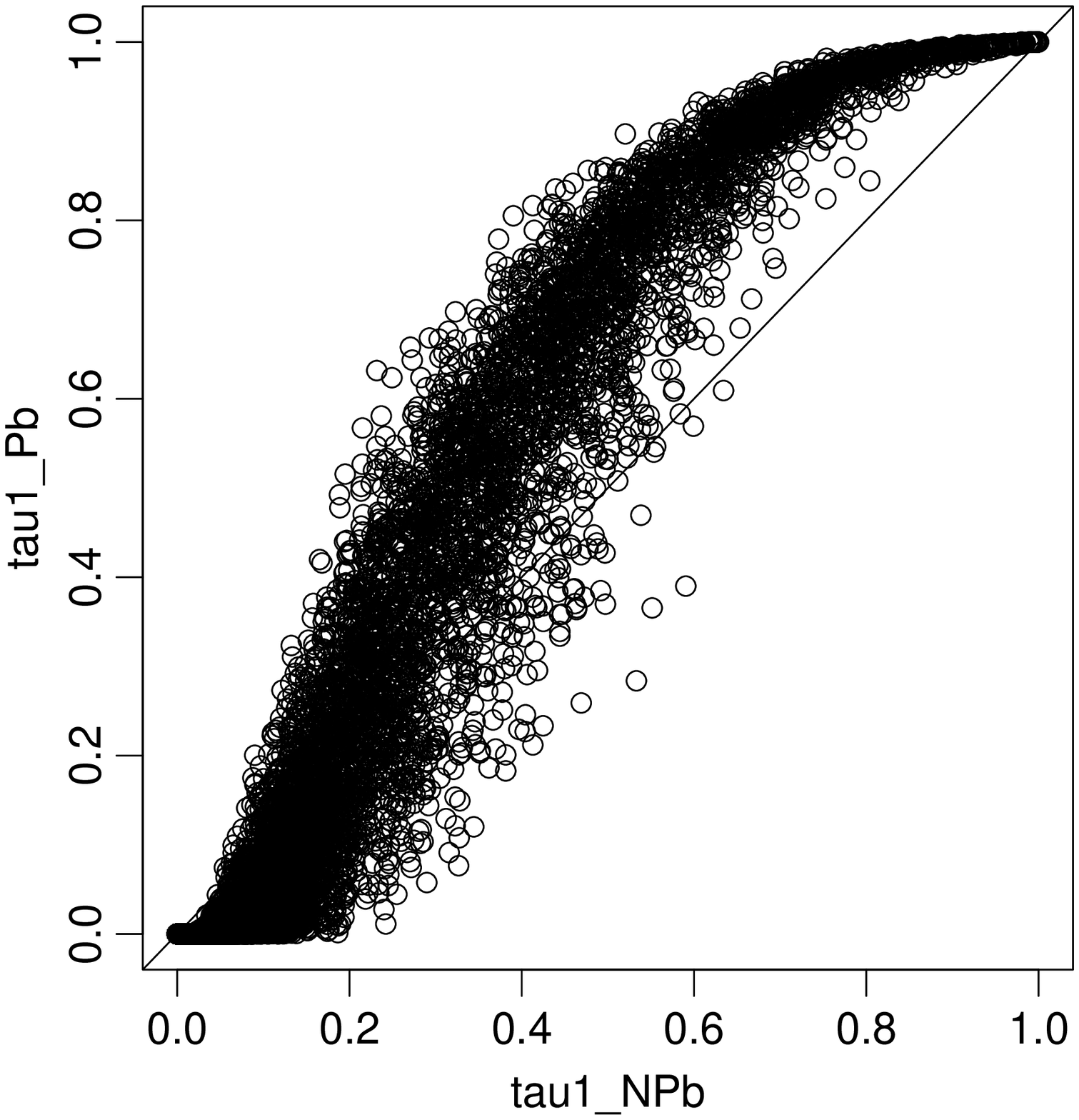} &
   \includegraphics[height=0.3\textwidth,angle=-90]{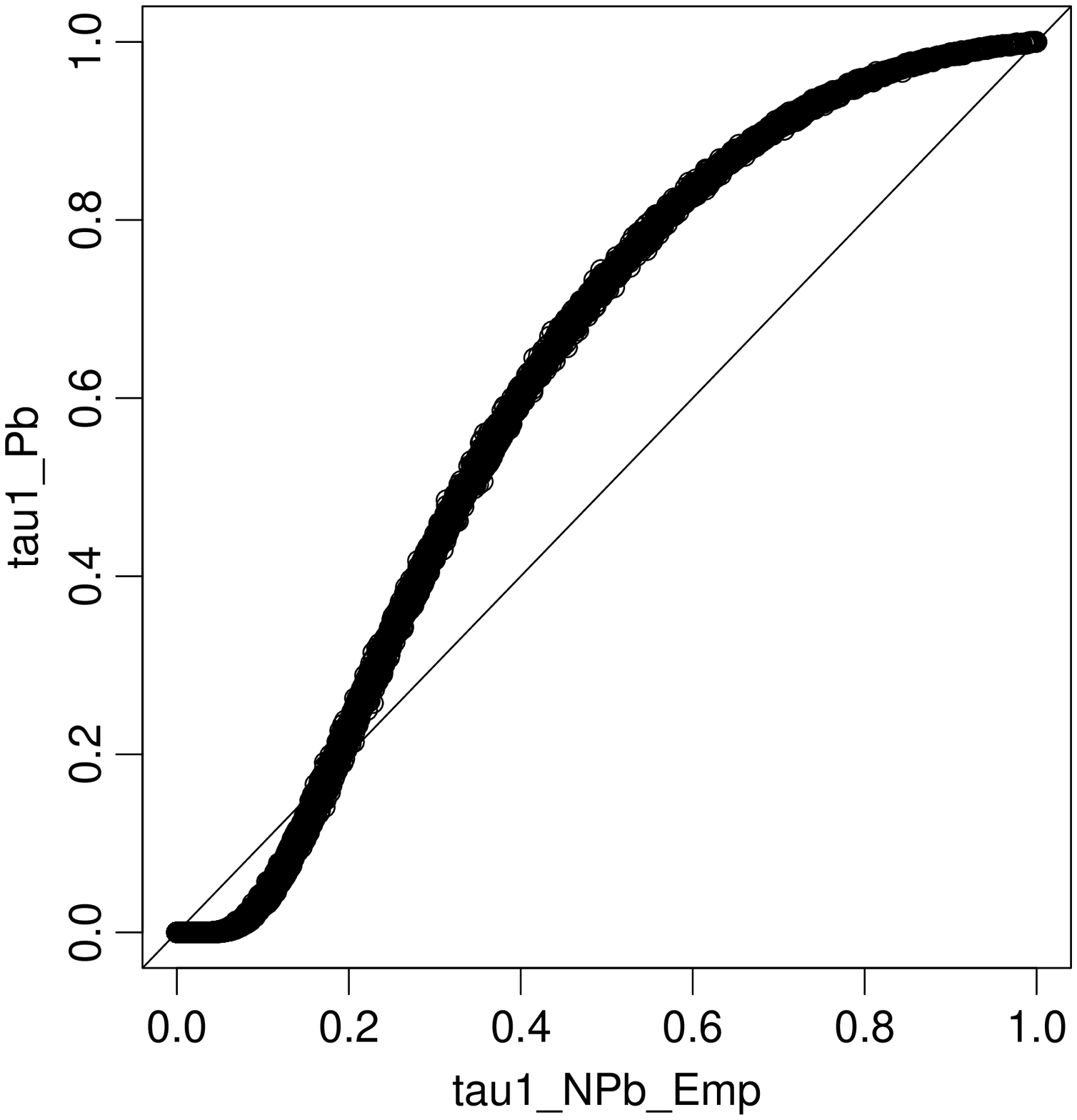}  
\end{tabular}
\caption{ {{\it California Large} table: nonparametric vs. parametric estimates of  ${\tau}_{1,k}$ in the presence of a vague Gaussian prior on ${\betavect}$ (column b), and under a degenerate prior  putting mass at the ML estimate of ${\betavect}$ (column c). The latter estimates are referred to as nonparametric empirical Bayesian estimates (labelled using the additional subscript ``Emp"). For completeness,  column (a) reports  nonparametric vs. nonparametric empirical Bayesian per cell risk estimates.}}
\label{fig:sigmoidinuovi:tot}
\end{figure}

Analysing per-cell risks, we next clarify how the over-estimation systematically associated with the ``too simple'' parametric models reported in Figure~\ref{fig:quantili:tab1e2} 
 is corrected for by the DP random effects at the cell level.
When comparing parametric vs. nonparametric estimates of per cell risks $\tau_{1,k}$ and $\tau_{2,k}$,  in all of our preliminary tests as well as under the models analysed in this paper we always noticed a typical sigmoidal shape shown  e.g. in the center and right-hand columns of Figure~\ref{fig:sigmoidinuovi:tot}, where estimates of $\tau_{1,k}$ are reported for the three models explored in the Large California table. (See  also Fig.~\ref{fig:sigmoidinuovi:tab1} in the supplementary material A.2, where a similar analysis is conducted focusing on the independence models considered in the  remaining three tables). This shape reveals that the presence of DP random effects invariably increases per-cell risk estimates that are small under the parametric model, but, even more, decreases per-cell risk estimates that are large under the parametric model, which  is the feature that invariably results in a significantly improved balance, i.e. reduced bias, at global level. 

The regularity and persistence of such adjustment 
  is impressive if one considers the diversity of the four contingency tables we are dealing with.  Indeed risks are estimated over tables of different sizes, with and without  structural zeroes, with very different proportions of population uniques, doubles and so on.  Notice that under the approach of \citet{skinner:shlomo}  the complexity of the 
\newpage
\noindent  optimal model increases remarkably with $K$, as shown in Tables~5-7 on p.~999 of their article where we can also observe an evolution of the starting model  from the independence (I) to the all two way interactions (II) model.   \citet[][online supplement]{mvreiter:jasa12} select the best log-linear  model according 
 to the criterion of  \citet{skinner:shlomo} for the  California Large table. Their results over samples of 5000 and 10000 individuals  confirm  that the complexity of the starting as well as the selected model strongly depends on the table's size.
     In contrast, under our approach, the increase of complexity needed to obtain very good fitting nonparametric models is extraordinarily limited. Moreover, the NP+I model invariably emerges as the starting model 
  \citep[see Figures~\ref{fig:quantili:tab1e2} and~\ref{fig:whipmidtau1quantiles}, and  compare with][where such model is presented as a sort of default model]{cflp2015}. All these findings not only induce us to conclude that the DP random effects are a formidable adaptive correction for the over-estimation found by \citet{skinner:shlomo} under ``too simple'', underfitting, log-linear models of type  (\ref{formula2}), but also invite us to take advantage of this by proposing a new  method for model selection.
        
     So far, we discussed a notion of positive bias due to underfitting and  arising in the model estimation phase; in Section~\ref{sec:tab:piccole1} we will elaborate on the negative bias due to overfitting. In the next section, instead,  we present our model selection procedure, which is derived with special attention to the different  sources of bias arising in the model selection phase  briefly reviewed in the initial paragraphs. We stress the difference with the criterion $\hat{B}$ of  \citet{skinner:shlomo}, that, as recalled in Section~\ref{intro1}, balances positive and negative bias arising in model fitting.
            \section{ New model selection method} 
           \label{sec:modsel}
          \citet{vehtariOj} give a comprehensive survey of established and recent Bayesian predictive methods for model assessment, selection and comparison.  
          Here we selectively review only those references useful to illustrate and comparatively discuss our proposal. \\
            The  literature \citep[see also][]{GelmanHV,US16,vehtariP} clearly indicates that the challenge in estimating predictive model accuracy is twofold: \\
         (i) to correct for the bias inherent in evaluating a model's predictions of the data that were used to fit it (within-sample-error), and \\ (ii) to address in some way the selection-induced  bias, i.e. the  undesirable optimistic bias in predictive performance evaluation  that, in  large sets of models, often leads to select a model by chance rather than by merit. 

       As to (i), several proposals of  bias correction are available in the literature \citep[discussed, for instance, in][and related articles]{vehtariOj,  GelmanHV, US16}, but they often incur in the second type of bias. As to (ii), it has long been known that any  criterion  suffers from selection bias \citep[see e.g.][and references therein]{zucchinilibro, Miller90,chatfield95, zucchini2000, vehtariOj,  GelmanHV, vehtariP}, and that it stems from a form of overfitting in model selection, analogous to the more familiar one occurring in training the model. In the last decade, however, the severity of this problem has  been re-affirmed and quantified for a series of established and recent criteria, especially in the machine learning literature \citep[e.g.][]{Reunanen03,VarmaS, CawleyT07,CawleyT10} where reliable model performance evaluation not only is crucial in many practical applications, but is required for fair comparison of machine learning algorithms.   A criterion with non-negligible variance has the potential for overfitting in the optimization phase by exploiting meaningless peculiarities of the sample over which it is evaluated.   Cross validation, widely used  criteria adjusting for within-sample-error such as AIC, DIC and WAIC which are approximations to different versions of cross validation, and many other criteria are all severely prone to selection bias \citep[see, e.g., ][]{ zucchini2000, CawleyT10, vehtariP}.  About (i) and (ii) and the underlying decomposition of the estimation error into bias and variance,  \citet{vehtariP}, p. 718, wrote: ``[...]the unbiasedness is intrinsecally unimportant for a model selection criterion'' and  ``it is more important to be able to rank competing models in an approximately correct order with a low variability.''  They also comment that, nonetheless, most literature focuses on unbiased estimates of the model predictive accuracy  and provides little guidance on how to reduce the selection induced bias. In our application, however, their solution, namely the projection approach in a so-called M-completed view  \citep[for details see][section 2.4]{vehtariP},  cannot be easily implemented, so we are left with traditional remedies against the  selection bias: restricting selection to a small number of well-considered models (this includes regularization and/or early stopping), or, alternatively, model averaging. \\ In what follows we propose a new model selection procedure and argue that  all of these remedies are in some way applied under our approach,  under the assumption that the true model is not necessarily included in the explored model space and that a good  model is just a useful approximation to the true model. As argued in \citet[][p.1006]{US16},  once one accepts this assumption, the focus immediately shifts to identifying which aspects of the model performance are most important to the end user, thereby relegating all the others to an intermediate, instrumental role.\\  

Our new two-stage model selection procedure is based on the so called ``score + search'' paradigm. The first stage is devoted to identify a path of search, i.e. which nonparametric models and in which order  have to be evaluated; at the second stage, an optimal model is selected through a new measure of model predictive accuracy specifically tailored to disclosure risk estimation.  The detail of the procedure is as follows.
       Building on findings in Section 2, we start from the nonparametric independence model, NP+I. At the first stage we focus exclusively on its parametric component, i.e. we restrict the search to fixed effects models of type (\ref{formula2}), and do a preliminary stepwise search.  
         Starting from a large value of a penalty factor $\gamma$, we gradually move it down and select, at each step,  
          the interaction term which maximizes a penalized log-likelihood, referred to as the criterion $C_0(\gamma)$,\\
                         \begin{equation}C_0(\gamma)=\sum_{k=1}^K log(p(f_k|\hat{\betavect}_{ML})) - d\times \gamma ,\label{C0}\end{equation}
             where $ p(f_k|\hat{\betavect}_{ML})= \frac{\pi^{f_k}}{f_k!} e^{ f_k\wvect'_k \hat{\betavect}_{ML} } e^{-\pi e^{ \wvect'_k \hat{\betavect}_{ML}}},$ $d$ is the difference between the number of parameters estimated under the current  model and under the  independence model I, and  $\gamma$ controls the strength of the penalty.  In principle this search is restricted to decomposable models (to have a  guarantee of existence of the ML estimates $ \hat{\betavect}_{ML}$, reliability of the degrees of freedom, and so on), but we will see that such a restriction is ineffective in practice, since we can stop the search after a few steps.\footnote{ At this  stage we  use  $ \hat{\betavect}_{ML}$ as an approximation to  Bayesian estimates of  the fixed effects $\betavect$ since they are assigned a vague prior (details in the supplementary materials A.1).}  \\
At the second stage of the procedure, a small set of candidate nonparametric models -  obtained by adding DP random effects to the parametric component identified at the first stage -   is evaluated through the application-specific criterion $C_1,$   \begin{equation}C_1=\sum_{k=1} ^K I(f_k=1) \times log\Big ( \int p(f_k|\lambda_k) p(\lambda_k|f_1,..,f_k) d\lambda_k  \Big),\label{C1} \end{equation}
where $p(f_k|\lambda_k)=\frac{1}{f_k!} {(\pi \lambda_k)^{f_k} e^{-(\pi \lambda_k)}}$ and $\lambda_k$ is defined as in (\ref{formula3}).  This is the {\it log pointwise predictive density} \citep[lppd, see][]{GelmanHV}  restricted to the unique cells, namely those crucial for estimating the global risks (\ref{eq:tau1}) and (\ref{eq:tau2}).  $C_1$  measures the model predictive accuracy, or performance, and is computed using posterior simulations $\lambdavect^{(h)}$, $h=1,..,H$:  $$ \sum_{k=1} ^K I(f_k=1) \times log\Big( \frac{1}{H} \sum_{h=1}^H p(f_k|\lambda_k^{(h)})\Big)$$  (see the supplementary material B for implementation details).  Following  \cite{US16}, and their description  of  different approaches to utility based model selection, $C_1$ can also be presented as a score based Bayesian information criterion whose particular scoring rule avoids that the good performance of a model on the subset of cells of interest is disguised by poor performance on the other cells, as it might be the case for criteria concerned with the model's performance across the full joint distribution. This is particularly appropriate in the case of disclosure risk estimation, as sample uniques usually are a very small subset of the total number of cells.   \\  Of course, models with high values of $C_1$ are preferred. Being high predictive accuracy equivalent to low predictive error, in this respect $C_1$ and the criterion $\hat{B}$ by \citet{skinner:shlomo} are not different.
Values of $C_1$ for  all models explored in   Figures~\ref{fig:quantili:tab1e2} and ~\ref{fig:whipmidtau1quantiles} (see also Table~\ref{tab:ModSETS} in the supplementary material A.1) 
are presented in 
Table~\ref{tab:C1largeCal} (computational details are provided in the supplementary material B).  In parentheses we show different models' rankings: based on $C_1$; on the true estimation errors, that is, the distance between the true value of $\tau_i$ and its  Bayesian point estimate under the model;  
  and based on the widely applicable information criterion 
 \citep{Watanabe09} restricted to the sample uniques, say $\text{WAIC}_\text{U}$. Although  parametric models are not candidate models, but just parametric counterparts of the candidate nonparametric models selected at the first stage of the procedure,  they are included in the rankings to show that 
\newpage
\begin{table}
\caption{ 
Risk estimates under the models 
explored in Figures~\ref{fig:quantili:tab1e2} and ~\ref{fig:whipmidtau1quantiles},  predictive measures and models' ranks (in brackets) based on the true estimation error,  on  $C_1$  and  $\text{WAIC}_\text{U}$. For each real data table we report the number of sample uniques (U) and the true values of $\tau_1$ and $\tau_2$.  
}.  \label{tab:C1largeCal}
{\footnotesize
\begin{tabular}{lcccc} 
\hline
Model & $\hat{\tau}_1$ & $\hat{\tau}_2$ & $C_1$ &  $\text{WAIC}_\text{U}$ \\
\hline
$California\, Large$ (U=11421) & ${\tau}_1= 2205$  &  ${\tau}_2= 3949.7$ & &\\ 
\hline
NP$_a$            & 2245.2 {\tiny{(1)}} & 4022.5 {\tiny{(1)}} & -21914.5 {\tiny{(1)}} & -30608.8 {\tiny{(5)}}\\ 
NP$_b$            & 2323.5 {\tiny{(3)}} & 4090.5 {\tiny{(3)}} & -22010.4 {\tiny{(2)}}  & -30676.1 {\tiny{(6)}}\\ 
NP$_c$            & 2272.2 {\tiny{(2)}} & 4034.3 {\tiny{(2)}} & -22032.9 {\tiny{(3)}} & -30199.5 {\tiny{(4)}}\\
P$_a$         & 2706.6 {\tiny{(5)}} & 4431.9 {\tiny{(5)}} & -29745.4  {\tiny{(6)}}& -29773.4  {\tiny{(3)}}\\
P$_b$         & 2727.1 {\tiny{(6)}} & 4458.4 {\tiny{(6)}} & -29594.9   {\tiny{(5)}}&-29631.1  {\tiny{(2)}} \\ 
P$_c$         & 2652.9 {\tiny{(4)}} & 4374.4 {\tiny{(4)}} & -29299.7  {\tiny{(4)}}& -29336.2  {\tiny{(1)}} \\ 
\hline
$California \, Medium$ (U=7669) & ${\tau}_1= 1169$  &  ${\tau}_2= 2314.6 $&  \\ 
\hline
NP$_a$            & 1185.4 {\tiny{(1)}} & 2340.6 {\tiny{(1)}} & -13624.5 {\tiny{(1)}}  & -18215.9 {\tiny{(4)}}\\
NP$_b$            & 1223.0 {\tiny{(3)}} & 2371.8 {\tiny{(3)}} & -13805.9 {\tiny{(3)}} & -18133.6 {\tiny{(2)}}\\ 
NP$_c$            & 1225.0 {\tiny{(4)}} & 2373.4 {\tiny{(4)}} & -13812.5 {\tiny{(4)}}  &-18098.1 {\tiny{(1)}}\\ 
NP+I     & 1189.2 {\tiny{(2)}} & 2345.3 {\tiny{(2)}} & -13632.6 {\tiny{(2)}} &-18195.9 {\tiny{(3)}}\\ 
P$_a$         & 1424.8 {\tiny{(8)}} & 2523.6 {\tiny{(8)}} & -18706.7   {\tiny{(8)}} & -18734.6  {\tiny{(8)}}\\ 
P$_b$         &1386.2 {\tiny{(5)}} & 2487.3 {\tiny{(5)}} & -18352.6  {\tiny{(5)}} &-18387.9  {\tiny{(5)}}\\
P$_c$         & 1399.4 {\tiny{(6)}} & 2500.9 {\tiny{(6)}} & -18364.4  {\tiny{(6)}} & -18399.1  {\tiny{(6)}}\\
P+I   & 1415.2 {\tiny{(7)}} & 2511.7 {\tiny{(7)}} & -18644.3   {\tiny{(7)}} & -18670.4  {\tiny{(7)}}\\ 
\hline
$California\,\, Small$ (U=3575) & ${\tau}_1= 498$  &  ${\tau}_2= 1023.4$ & \\ 
\hline
NP$_a$            & 479.8 {\tiny{(3)}} & 1003.2 {\tiny{(3)}} & -6212.9 {\tiny{(3)}} &-7886.3 {\tiny{(1)}} \\ 
NP$_b$            & 483.8 {\tiny{(1)}} & 1011.3 {\tiny{(1)}} & -6183.8 {\tiny{(2)}}  & -7993.0 {\tiny{(3)}}\\
NP+I     & 480.3 {\tiny{(2)}} & 1008.6 {\tiny{(2)}} & -6175.9 {\tiny{(1)}}&-7982.6 {\tiny{(2)}}\\ 
P$_a$         &581.6 {\tiny{(6)}} & 1072.9 {\tiny{(5)}} & -8951.8  {\tiny{(4)}}  & -8972.9  {\tiny{(4)}} \\ 
P$_b$         & 568.5 {\tiny{(4)}} & 1065.0 {\tiny{(4)}} & -9023.4   {\tiny{(5)}} & -9055.2  {\tiny{(5)}}\\ 
P+I   & 579.9 {\tiny{(5)}} & 1077.6 {\tiny{(6)}} & -9109.1   {\tiny{(6)}} &-9131.4  {\tiny{(6)}}\\
\hline
$WHIP$ (U=7176) & ${\tau}_1= 915$  &  ${\tau}_2= 1948.1$ &  \\ 
\hline
NP$_a$            &~917.9 {\tiny{(1)}} & 1981.2 {\tiny{(3)}} & -12022.0 {\tiny{(2)}}& -16107.4  {\tiny{(5)}}\\ 
NP$_b$            & 1003.1 {\tiny{(5)}} &2078.4 {\tiny{(5)}} & -12261.6 {\tiny{(5)}}  & -16413.3 {\tiny{(7)}}\\ 
NP$_c$            & ~921.2 {\tiny{(3)}} & 1987.0 {\tiny{(4)}} & -12128.4 {\tiny{(3)}} & -15977.5 {\tiny{(3)}}\\ 
NP$_d$            & ~908.9 {\tiny{(2)}} & 1972.2 {\tiny{(2)}} & -12134.7 {\tiny{(4)}}& -15767.7  {\tiny{(2)}}\\%
NP$_e$            & ~874.8 {\tiny{(4)}} & 1930.2 {\tiny{(1)}} & -12010.1 {\tiny{(1)}} &-16084.5 {\tiny{(4)}}\\ 
NP+I       & 1010.4 {\tiny{(6)}} & 2083.4 {\tiny{(6)}} & -12149.9   {\tiny{(5)}}& -16195.7  {\tiny{(6)}}\\ 
P+I         &1184.9 {\tiny{(7)}} & 2289.9 {\tiny{(7)}} & -15633.6 {\tiny{(7)}} & -15650.3   {\tiny{(1)}}\\ 
\hline
\end{tabular}
}
\end{table}
 \noindent
 in some contingency tables (Large California and WHIP) the $\text{WAIC}_\text{U}$  selects a  largely sub-optimal parametric model, despite the small number of models under evaluation. This is  the empirical evidence of substantial selection bias and optimism in the performance evaluation due to the increase of the variance of the criterion implied by the presence of a further estimated term.\footnote{To be more precise,  $\text{WAIC}_\text{U}$ is obtained by adding to  $C_1$  a data based  bias correction which,   analogously to $p_{WAIC2}$ in  \citet{GelmanHV}, uses the posterior variance of  individual terms  in the log-predictive density summed over the $U$ sample uniques.} Instead, $C_1$ provides a reasonably good ranking of models  in all contingency tables, and   when (see, for instance, second and third positions in the Large California table and second and first positions in the  Small California table)  the rankings based on the true estimation error do not agree with those based on $C_1$, yet the relative positions defined by $C_1$ do not differ substantially from the ``true'' ones. Importantly, moreover,  the corresponding values of $C_1$ are  so close to each other that we are actually  warned about  possible inversions of positions  in the ranking.
In the WHIP table it is also worth noticing that the NP$_e$ and $NP_d$ models, featuring four interaction terms, turn out to be  good models essentially because of the presence  of an interaction term (ESEC*WORKP, as can be seen in  the supplementary material A.1, Table~\ref{tab:ModSETS}) already  selected  at the first stage  of the procedure by means of  $C_0(\gamma)$ and included in two models,  NP$_a$ and NP$_c$, among which $C_1$ selects the one including this two-way interaction only. It is this term that marks the difference between good and inadequate models; in fact, in all our preliminary explorations  we observed that adding to the independence model a single interaction term (selected by using $C_0(\gamma)$) is enough to enter the range of reasonably good nonparametric models. A theoretical justification to this is provided later, within the discussion of our model selection method. In the remainder of the section we will further comment on  the  pair ($C_0$,  $C_1$)  and the results they produce with special attention to both the challenging issues (i) and (ii) recalled at the beginning of the Section. Furthermore, we will discuss the proposed procedure in light 
 of the specialized literature on model selection for disclosure risk estimation reviewed in Section~\ref{intro1}.  We will see that the problem of model choice, which is a very difficult task in a strictly parametric context, is significantly defused by our approach. 
 
 As regards (i), 
 as  already stressed,  the criterion $C_1$ is based solely on the sample uniques, i.e. a very small subset of the $K$ cells 
 used to fit the model.  For illustration, 
 see  Table~\ref{tab:C1largeCal} where  $\text{U}/K\leq 0.001$ for all California tables and $\text{U}/(K-\text{\# structural zeroes})\leq 0.0169$ for the WHIP table. This fact results in a negligible within-sample-error leading us to omit a bias correction term. Importantly, this means that no additional variability is introduced in the criterion by the bias adjustement.  A  comment of the same nature can be found in \citet[][p.53]{zucchini2000} about the {\it  simple} criterion $C^*_{K-L} $ compared to its bias corrected version  $C_{K-L} $. With reference to  \citet[][p.1027]{US16},   whose BPSIC  is a Bayesian, very adaptable and refined evolution of the criterion  $C_{K-L}$  \cite[see also][]{zucchinilibro}, our omission can be interpreted as an extremization  of the general benefit represented by the lower bias correction term applicable when a criterion is ``based on the relevant  marginal and  conditional logarithmic scores of the variables of interest within a larger model''. \\  
As regards  (ii), we first comment on $C_0(\gamma)$, the criterion used to select suitable candidate nonparametric models at the first stage of the procedure. While based on a double use of the $K$ sample frequencies, in place of a bias correction term limiting the within-sample-error, (\ref{C0}) includes a relevant ``economic'' component, $d \times \gamma$, due to  the large values of $\gamma$  we employ in the search. This is a request of simplicity, independent of the sample, directly suggested by the great ability of the DP random effects to correct for the over-estimation associated with under-fitting parametric  log-linear models observed in Section 2. As a matter of fact, this structure of $C_0(\gamma)$ limits two very different types of bias at the same time: the selection bias (avoiding the additional variability introduced by the bias adjustment) on the one hand, and the unbalance of over-estimates and under-estimates of per cell risks under over-parametrized nonparametric models on the other, as  we will see in Section~\ref{sec:tab:piccole1}. Of course, such a structure of $C_0(\gamma)$ indirectly implies a strongly non uniform prior on the models under consideration (in principle all decomposable log-linear models). Finally, $C_0(\gamma)$ is not endowed with a stopping rule since it is used jointly with the application-specific criterion $C_1$. The search stops when, for nonparametric models of increasing complexity, $C_1$ ceases to improve and begins to decline.   For instance, re-using the expression of \citet{skinner:shlomo}, in all contingency tables  presented in Section 2 two two-way interactions identified through $C_0(\gamma)$ are enough to enter the range of ``reasonably good" nonparametric models according to $C_1$. This also means  that, in different applications,   $C_0(\gamma)$ can be employed to identify suitable nonparametric log-linear models jointly with different application-specific criteria. Comparatively, therefore, it emerges as a general purpose criterion.   Turning now to the second stage criterion $C_1$, we point out that each of the nonparametric models evaluated through (\ref{C1}), or, possibly, different application-specific criteria, is an average model. This can be seen in the likelihood $L(\betavect,m|f_1,\ldots,f_K) $,
 which is a sum of $B_K$ terms, where $B_K$ is the Bell number, resulting from all possible partitions (clusterings) $C$  of the $K$ sample frequencies   $f_1,\ldots,f_K$ in $c$  nonempty clusters ($1\leq c\leq K$) \citep[see, e.g.,][]{lo,liu}. Denoting by  $n_j$ the number of cells included in the j-th cluster ($1\leq  n_j \leq K$), 
 we can interpret  each term in this sum\footnote{in our case  \citep[see][]{cflp2015} it can be written as follows:   $L(\betavect,m|f_1,\ldots,f_K) =$ $$
\sum_{c=1}^K  \,\,\sum_{C:|C|=c}  \,\,\frac{\Gamma(m) \, m^c }{\Gamma (m+K)} \, \,\,\, \prod_{j=1}^c  \Gamma (n_j) \int {\prod_{k\in \mathrm{cluster}\,\,j\,\,} \frac{\pi^{f_k}}{f_k!} e^{ f_k(\wvect'_k \betavect + \phi_j)} e^{-\pi e^{(\wvect'_k \betavect +\phi_j)}}} 
dG_0(\phi_j), 
$$ 
. } as the product of two factors:  
\begin{equation}
\label{likpar}
 \prod_{j=1}^c \,\, \int {\prod_{k\in \mathrm{cluster}\,\,j\,\,} \frac{\pi^{f_k}}{f_k!} e^{ f_k(\wvect'_k \betavect + \phi_j)} e^{-\pi e^{(\wvect'_k \betavect +\phi_j)}}} 
dG_0(\phi_j),  
\end{equation} 
i.e.  the  likelihood corresponding to a {\it parametric} log-linear mixed model  with the same (very few) fixed effects and a $G_0$  distributed random effect  specific to each cluster $j$  belonging to  a fixed partition $C$ of $f_1,\ldots,f_K$  into $c$ clusters, times  $$[\Gamma (m+K)]^{-1} \, \Gamma(m)\,  m^c  \,  \Gamma (n_1) \times \cdots \times  \Gamma (n_c),$$  i.e. the probability  $Pr\{n_1,\ldots,n_c|C, c \}$  assigned to such partition by the multivariate Ewens distribution \citep[][chap. 41]{takemura,Johnson}. In other words,  $L(\betavect,m|\mathbf{f})$  is an average of $B_K$ {\it parametric} likelihoods (\ref{likpar}) according to specific  weights on  random partitions of the $K$ sample frequencies. 
 This implies various, useful consequences. Of special interest here is that, as $K$ increases, the stimulus received by the mechanism of model averaging  just described is extraordinarily strong, since the number of terms summed in the likelihood 
 increases as follows $$B_{K+1}=\sum_{s=0}^K \binom{K}{s} B_K. $$
This massive model averaging action implied by the presence of DP random effects strongly limits the need for additional interaction terms to obtain good models (see  Table~\ref{tab:C1largeCal}), and explains the permanence of the NP+I model as the starting model.  
More explicitly, each given candidate nonparametric model is extraordinarily boosted  by the increase in $K$, since, for each  partition of the $K$ sample frequencies, a possible relation of dependence among observations in the same cluster is explicitly evaluated and exploited for inference. As opposite, as $K$ increases, the parametric counterpart  of each candidate nonparametric model becomes more and more under-fitting.  This explains the increasing gap between performances of parametric and nonparametric models observed in  Figure~\ref{fig:quantili:tab1e2} for California tables of increasing sizes (roughly,  going from the Small to the Large table, the estimation error under  parametric models increases tenfold.)  In conclusion, the strongly adaptive reinforcement of each candidate nonparametric model that occurs under our approach to model selection  induces both a data-driven significant  restriction of the search space  and a reduction of the sensitivity of risk estimates to the specification of the model, that is a form of robustness. Such two facts concur to produce a substantial simplification of the model selection task.  \\ 
  In comparison with \citet{forster:webb},  in our model selection procedure model averaging and restriction of the set of possible specifications for fixed effects to decomposable structures  are applied in reverse order.  Further distinctive points of difference are that in our case: 1)  model averaging stems automatically from having extended the model class to the  family of log-linear mixed models with DP  random effects; 2)  for each candidate nonparametric model,  averaging is performed over $B_K$ parametric models according to the weights just discussed, rather than over the inferences corresponding to the whole class of graphical log-linear models according to weights given  by a posterior distribution on the whole  model space; 3) the restriction of the possible alternative specifications of fixed effects to the decomposable structures is  ineffective in practice, given the extreme closeness of reasonably good nonparametric  log-linear models to the starting model, NP+I. 
   This is indicated in all our examples by the values of $C_1$,  which induce to stop the search early, and is confirmed by benchmarking our estimates to the true values of risks. 
In comparison with \citet{skinner:shlomo}, our model selection procedure not only avoids possible  problems with  nonexistence of ML estimates of their candidates  of type (\ref{formula2}), but also  limits the selection-induced bias due to the large number of models under evaluation. 
Finally, a point by point  comparison with \citet{mvreiter:jasa12} is less agile because of the very  different modeling framework  they introduce to describe contingency tables. Overall, however, our approach  to model selection for disclosure risk estimation relies on a pair of simple, easily applicable, criteria and such a strong reduction of the search space that, in comparison with standard practice, it probably is a way forward.\\
 

\section{Overfitting and under-estimation: ideas for small area estimation}\label{sec:tab:piccole1}
In this Section we contrast more and less parsimonious nonparametric models for risk estimation, and reconsider the bias that arises in the model fitting phase, 
 exploring models of  type (\ref{formula3}) in a different setting, namely in a  small and dense contingency table.  The reason of this choice is twofold: 
 from a practical  perspective, this exploration is a test whose results are useful in applications where the size and sparsity of tables are not  as extreme as  in disclosure risk estimation;  from a technical perspective, such choice guarantees against the severe issues arising when fitting complex log-linear models in the presence of sparse tables  \citep[e.g.][]{fienberg2012}. 
  
For illustration, we reconsider the WHIP data (the  7\% microdata sample from the Italian National Social Security Administration 2004).  As described in the supplementary material A.3, we now 
obtain a new contingency table of size $K=3,960$, referred to as the Small WHIP table, based on five spanning (key) variables.
With this data at hand, from within our enlarged class of models  (\ref{formula3}), we reconsider the association between under-estimation of risks and specification of overfitting models discussed in \citet{skinner:shlomo}.  
So far, the performance of the family of nonparametric log-linear mixed  models  with DP random effects has been evaluated with the purpose of estimating an overall measure of disclosure risk. Here, given 
that global risks are estimated by summing the cell-specific quantities $\hat \tau_{i,k}$, $i=1,2$, we perform a close analysis of the latter in order to draw  useful directions for future research.
In particular, to highlight that  nonparametric models with a small or  a large number of parameters may  complement each other when different viewpoints are taken, we  now compare two ``reference'' models,  namely  the nonparametric independence and all-two-way interactions models (NP+I and NP+II, representative of more and less parsimonious models)  with respect to both of the above recalled estimation goals, that is,  global and cell-specific risk estimation.
We show that a model with a large number of parameters  having a poor performance in  global disclosure risk estimation, i.e.  an  ``over-parametrised"/overfitting  model for this purpose, may perform better than a less parametrised model at estimating the per cell risks  over certain subsets of sample unique cells, specifically, the low risk cells.
 For illustration, it will also be useful to consider the fully parametric counterparts of these models (P+I and P+II, respectively).

\begin{figure}[hb]
 \centering
 \includegraphics[width=0.44\textwidth, angle=-90]{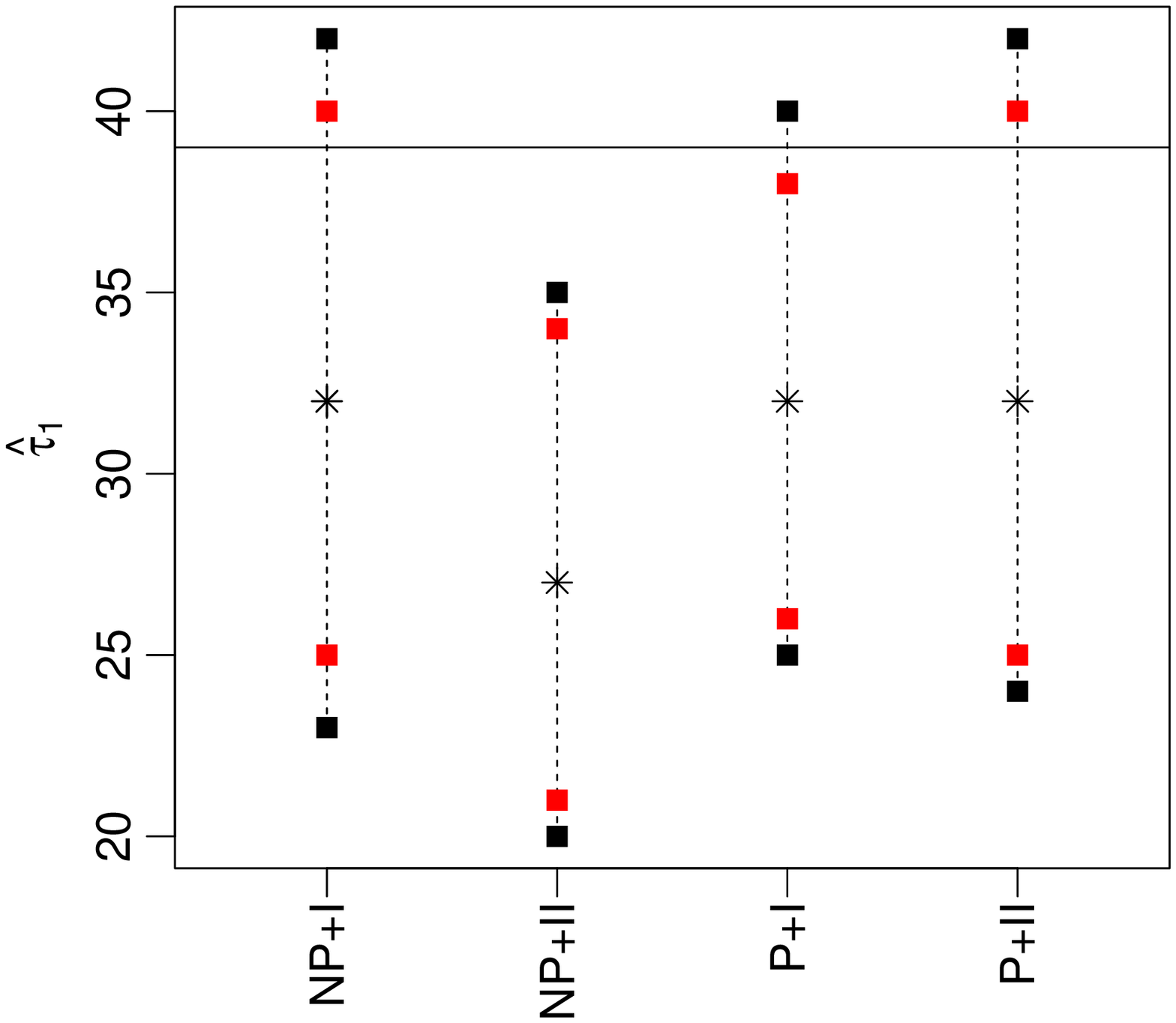}   
  \includegraphics[width=0.44\textwidth,  angle=-90]{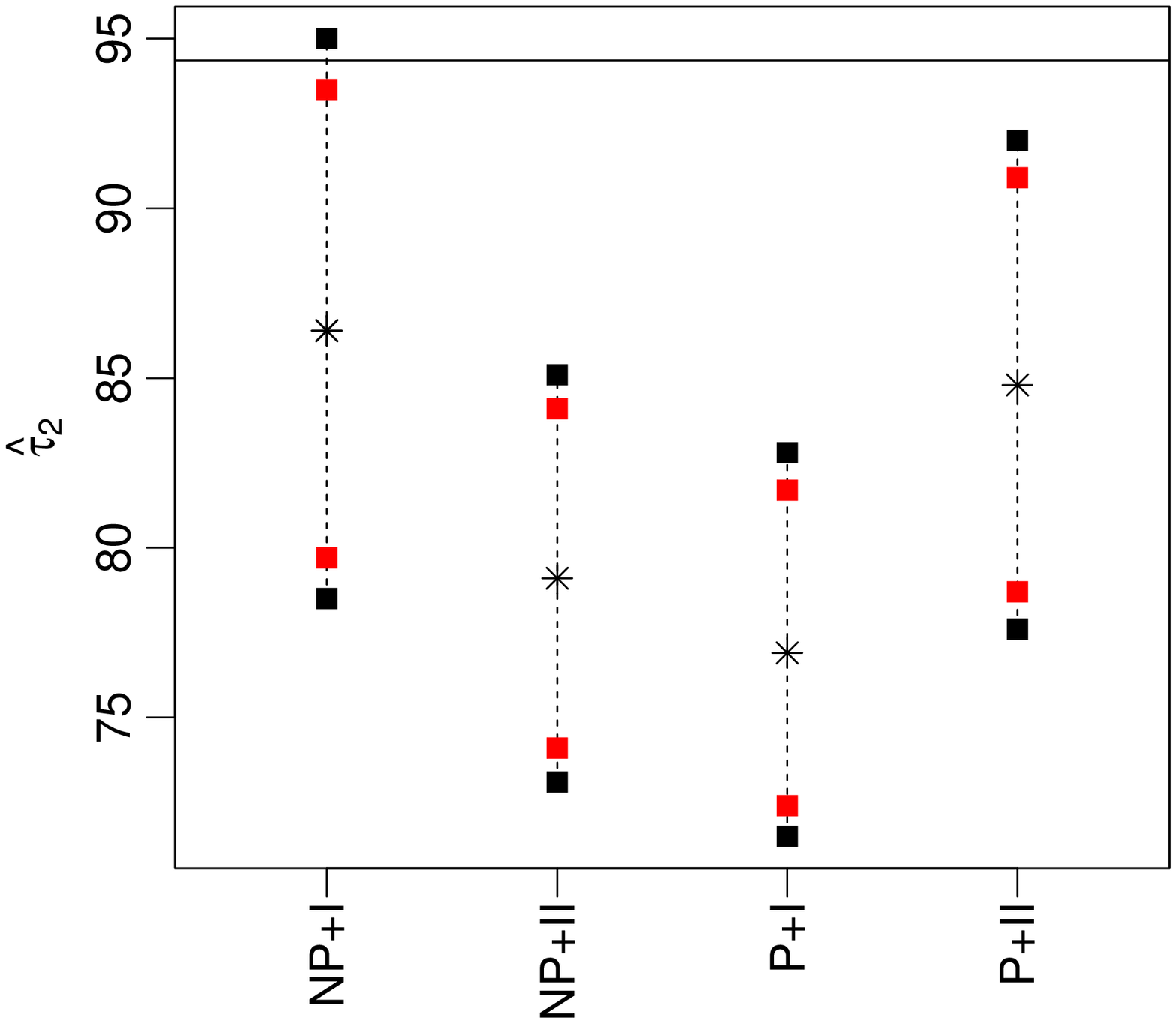}   
  \caption{True values (horizontal solid line) and quantiles  (0.005, 0.025, 0.50, 0.975, 0.995)  of the posterior distributions of ${\tau}_{1}$ (left) and ${\tau}_{2}$ (right) under all parametric and nonparametric models considered  for the Small WHIP table.  }
 \label{fig:quantili}
\vspace{\SPACE}
 \end{figure}

\begin{figure}[h]
 \centering
 \includegraphics[height=0.6\textwidth,angle=-90]{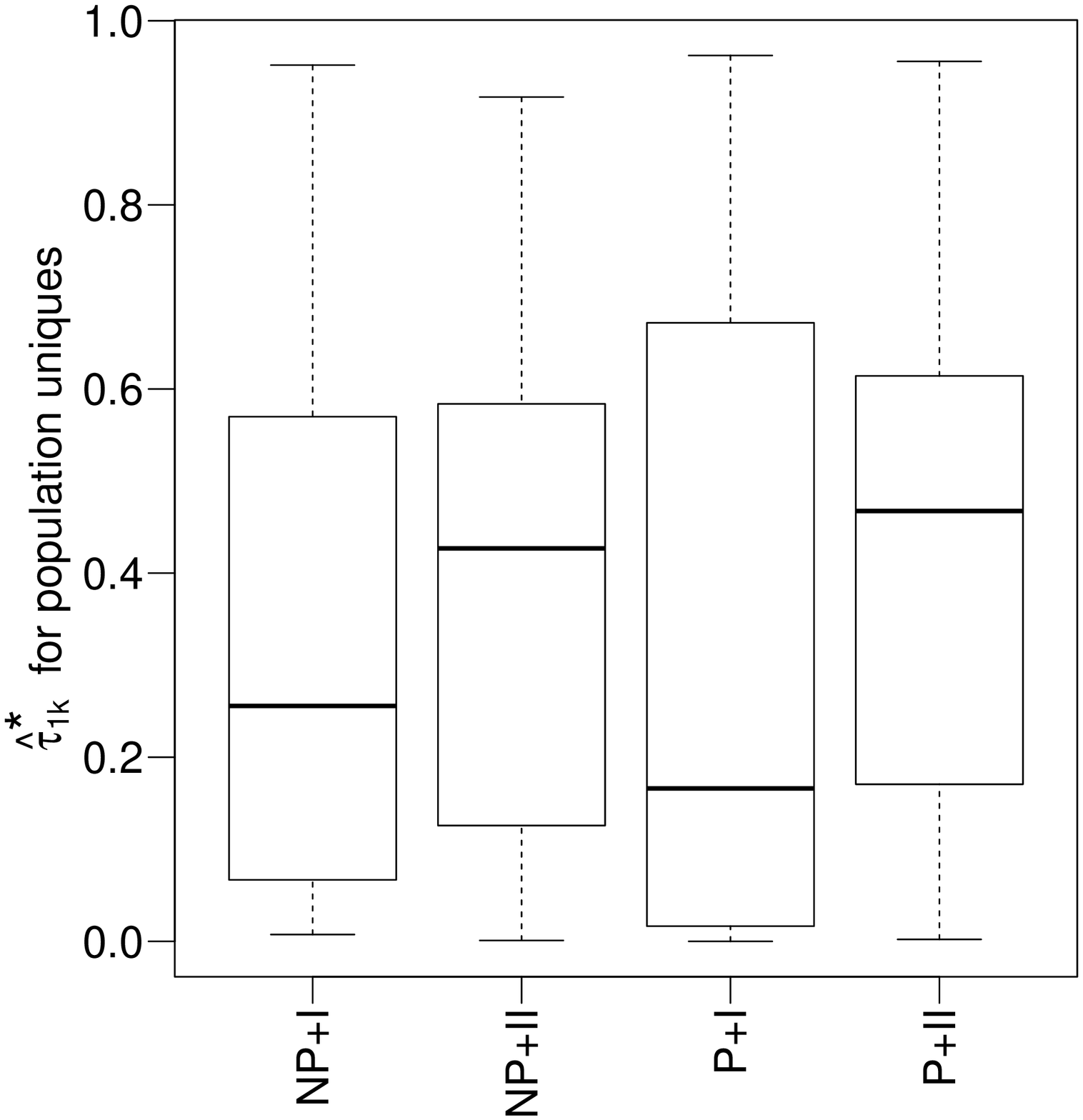} 
 \caption{Boxplots of the estimated values of  $\tau_{1,k}^*$ {\it{ for population uniques only, i.e. cells where $\tau_{1,k}=1$}}, under the nonparametric  independence and all two-way interactions  models, NP+I and NP+II, respectively, and their parametric counterparts, P+I and P+II;  Small WHIP table. }
 \label{boxplot:tau1k}
\vspace{\SPACE}
\end{figure}

\begin{figure}
   \centering
  \includegraphics[height=.4\textwidth,angle=0]{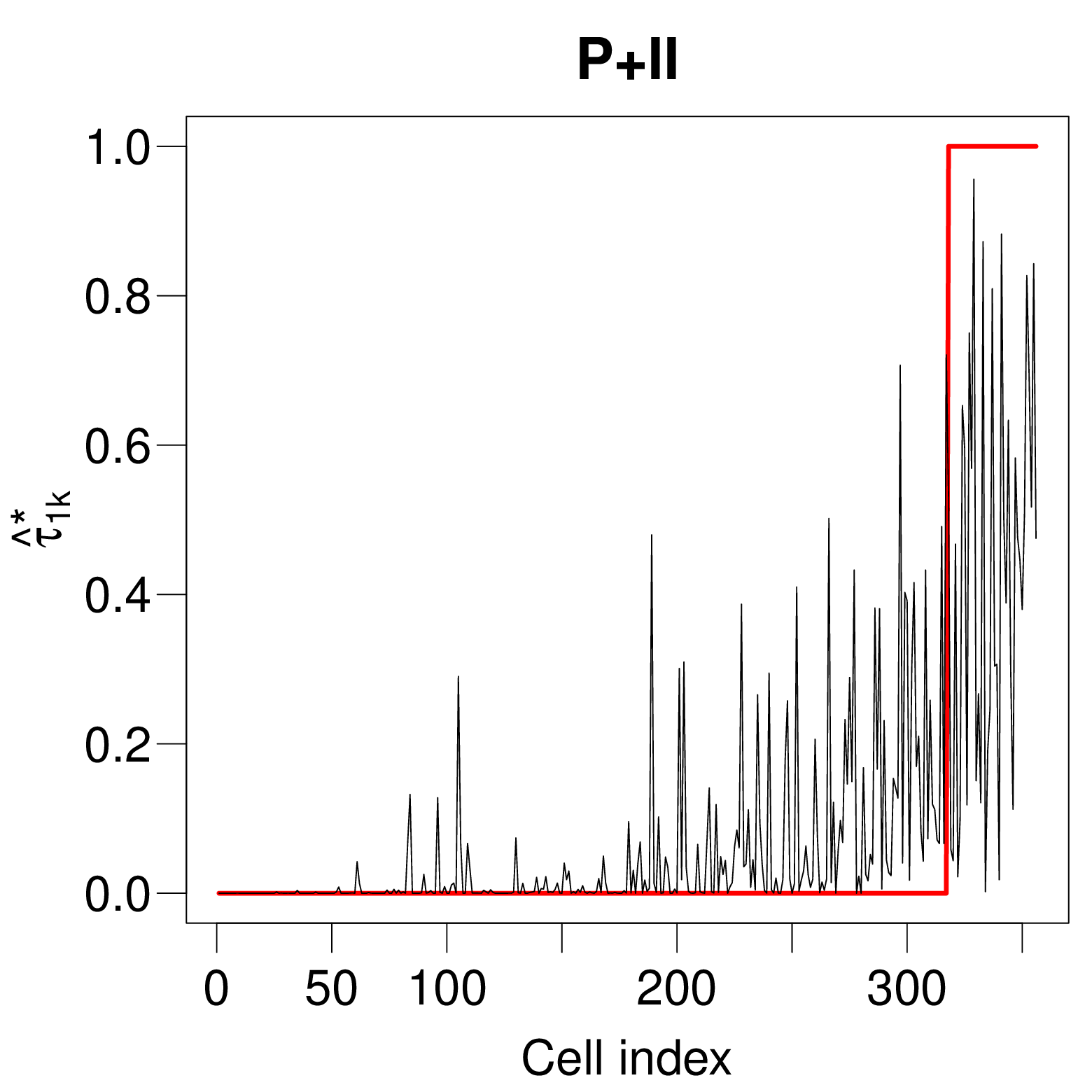}  
  \includegraphics[height=.4\textwidth,angle=0]{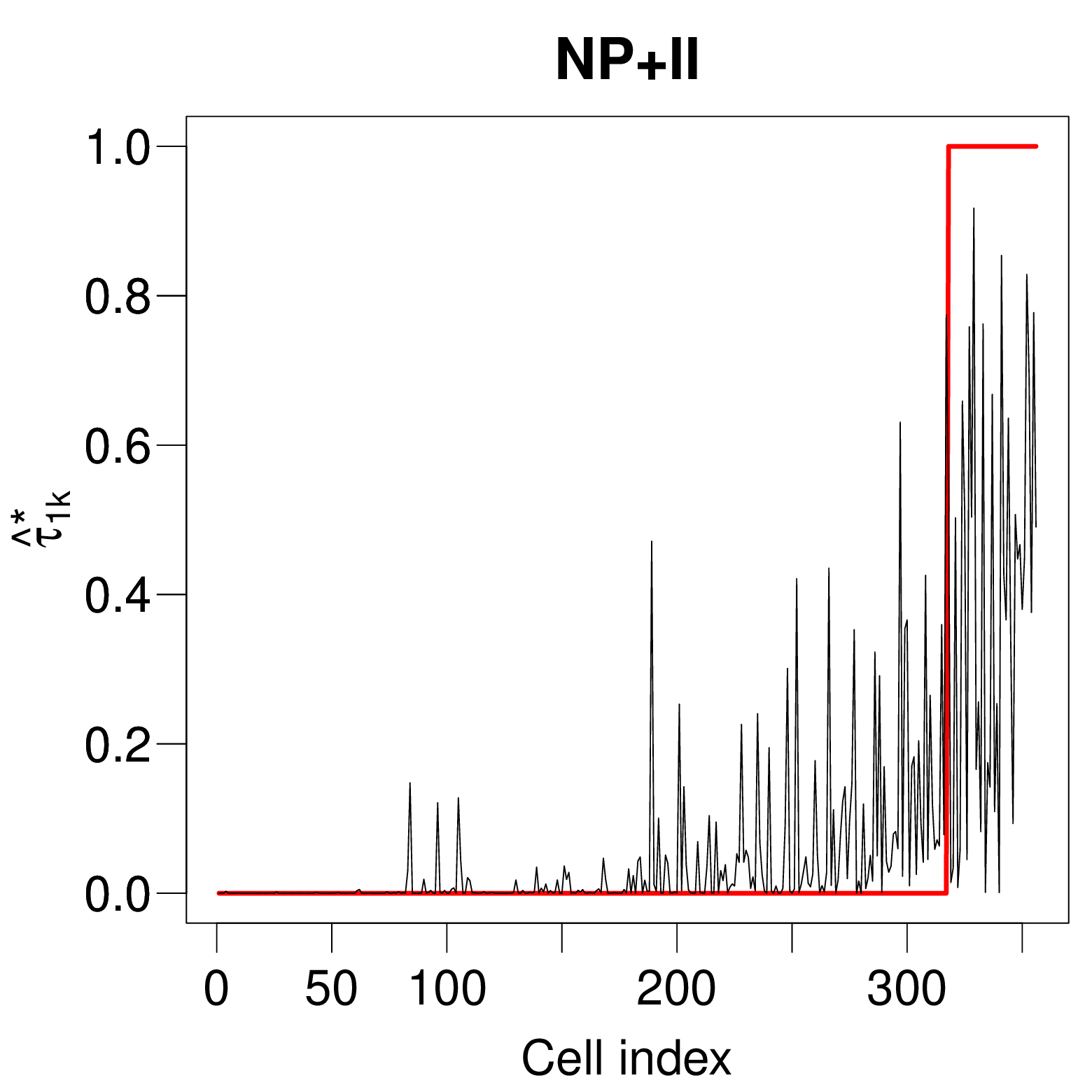} 
  \\
  \includegraphics[height=.4\textwidth,angle=0]{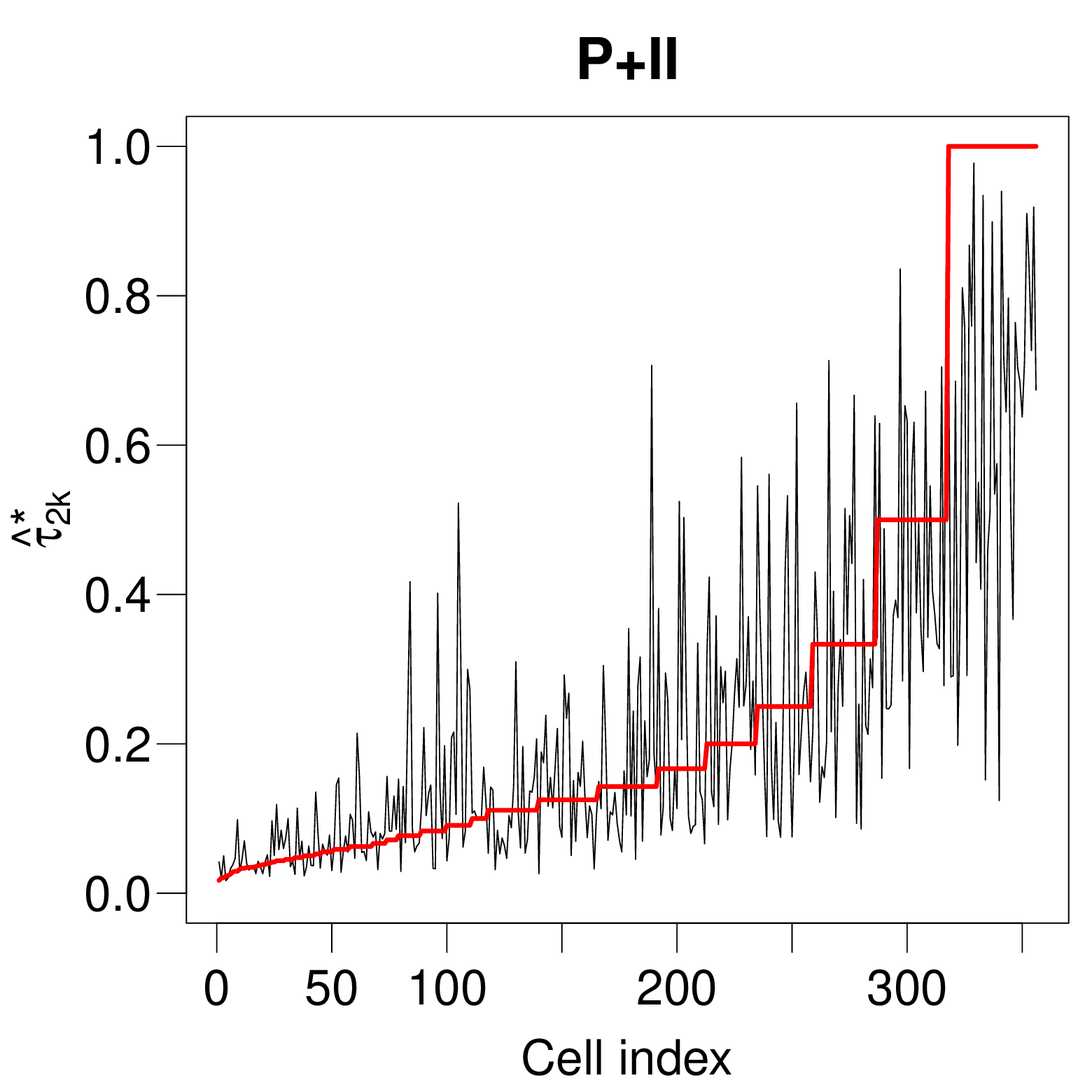}  
    \includegraphics[height=.4\textwidth,angle=0]{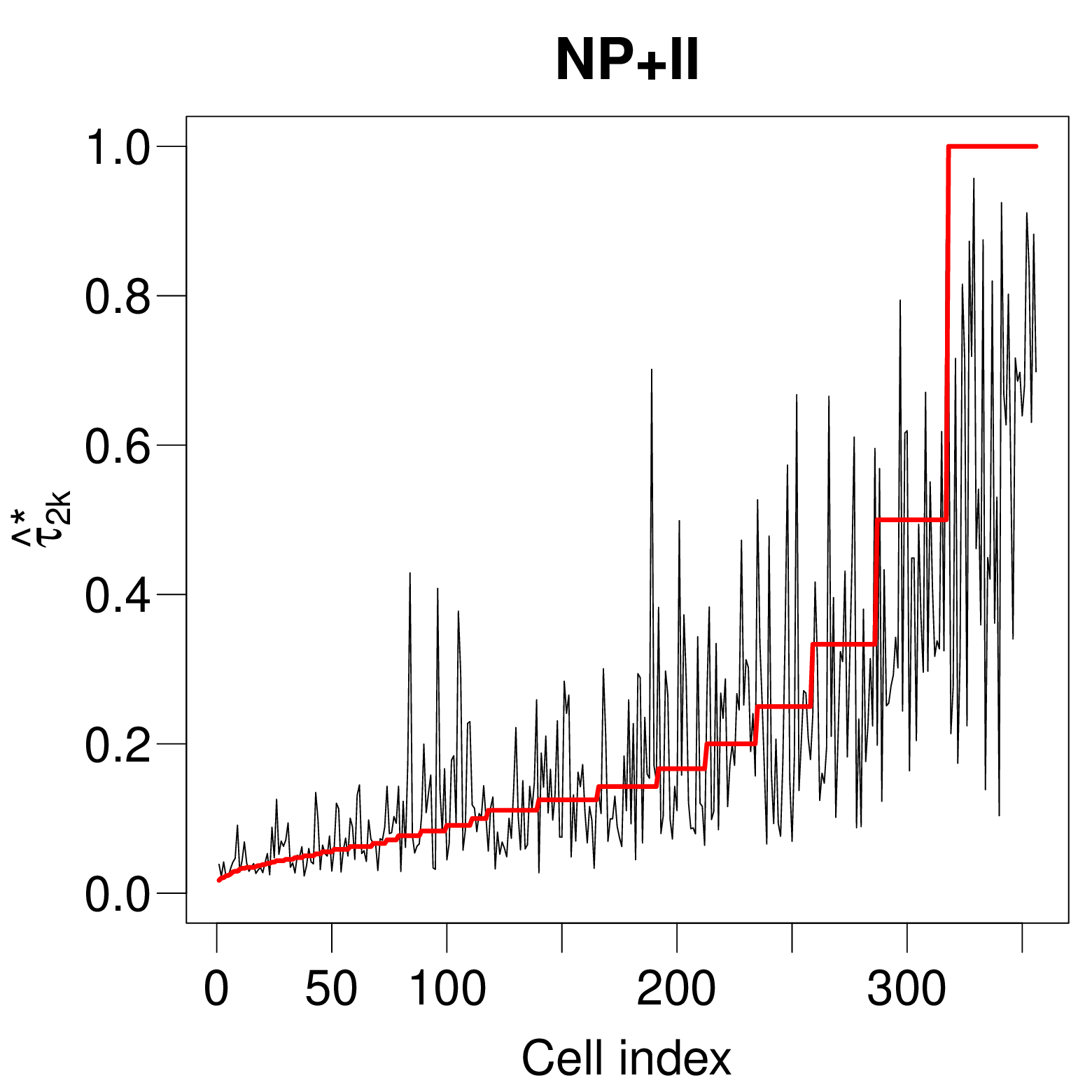} 
  \caption{Risk estimates $\hat{\tau}^*_{1,k}$ (top row)  and $\hat{\tau}^*_{2,k}$ (bottom row) for sample unique cells under the all-two-way interactions parametric (left) and nonparametric (right) models  in the Small WHIP table. Cells are arranged in increasing order of true per-cell risk (red line).}
  \label{fig:tau2k:IIvsII}
\vspace{\SPACE}
  \end{figure}

\begin{table}[tb]
\caption{ Estimated values of $\tau_1$ and $\tau_2$ by means of both estimators $\hat{\tau}^*_1$ and $\hat{\tau}_1$, and $\hat{\tau}^*_2$ and $\hat{\tau}_2$ (s.e. in parentheses)  for the Small WHIP table. True value of the global risks are 
$\tau_1 = 39$ and $\tau_2 = 94.4$. 
\label{tab:results3}}
\begin{tabular}{p{2.4cm}p{1.8cm}p{1.8cm}p{1.8cm}p{1.8cm}}
\hline

Model & $\hat{\tau}^*_1$ & $\hat{\tau}_1$ & $\hat{\tau}^*_2$ & $\hat{\tau}_2$ \\

\hline
NP+I        & {\bfseries{32.1}} {\tiny{(2.5)}}  & {\bfseries{32.1}} {\tiny{(4.7)}} & {\bfseries{86.5}} {\tiny{(3.2))}} &  {\bfseries{86.5}} {\tiny{(4.2)}} \\ 
NP+II       & {\bfseries{27.4}} {\tiny{(1.4)}} & {\bfseries{27.4}} {\tiny{(4.0)}} & {\bfseries{79.1}} {\tiny{(1.6)}} &   {\bfseries{79.1}} {\tiny{(3.1)}}   \\ 
P+I                & {\bfseries{32.1}} {\tiny{(1.0)}} & {\bfseries{32.1}} {\tiny{(3.9)}} &{\bfseries{76.9}} {\tiny{(1.4)}} & {\bfseries{77.0}} {\tiny{(2.8)}} \\
P+II               & {\bfseries{32.5}} {\tiny{(2.4)}}  & {\bfseries{32.5}} {\tiny{(4.6)}} &  {\bfseries{84.8}} {\tiny{(2.5)}} & {\bfseries{84.8}} {\tiny{(3.7)}} \\ 
\hline
\end{tabular}
\vspace{\SPACE}
\end{table}
 Table~\ref{tab:results3} reports true and  estimated values  of the global risks  $\tau_1$ and $\tau_2$ (standard errors, s.e., in parentheses). Plots in Figure~\ref{fig:quantili}  present the 2.5th, 5th, 50th, 95th and 97.5th percentiles  of the posterior distribution of $\tau_i$, i=1,2, under the same models. Details about priors on model's parameters and computations are in the supplementary materials A.3 and B, respectively. In  Figure~\ref{fig:quantili}  we notice the expectedly good performance of the default nonparametric model NP+I, roughly 
comparable to that of the 
parametric model, P+II. At the same time, we observe that for nonparametric models the bias in estimating global risks increases  with the number of fixed effects (see also  Table~\ref{tab:results3}).  
 Moreover,  Table~\ref{tab:results3} provides a clear indication that, as the model complexity 
 increases, the variability corresponding to parametric models increases as well, while the variability corresponding to nonparametric models tends to decrease.  
 Since all of these phenomena are clearly noticeable by considering $\hat{\tau}_i$ or  $\hat{\tau}^*_i,$ $i=1,2,$ to quickly get at the root of 
 the behaviour of our estimators, in the rest of the Section we
 focus on  per-cell risk estimates $\hat{\tau}_{i,k}^*$, $i=1,2$. This means that the variability of the $F_k$s is neglected,  but provides us with an effective simplification. For instance, 
 Table~\ref{tab:results3} clearly shows that the NP+II  model is an overparametrized  model when estimation of the global risks (\ref{eq:tau1}) and (\ref{eq:tau2}) is seeked; but let us  now consider  the estimation of per cell risks.
\begin{table}[h]
\caption{Signed, absolute and squared errors for the estimation of $\tau_{2k}^*$ under all models considered in the Small WHIP table: for all cells  (left panel), restricted to cells having large frequency in the population ($F_k>3$ and $ F_k>10$; central panel), and restricted to cells having  small  frequency in the population ($F_k\leq 3$ and $ F_k\leq 10$; right panel).}
\label{tab:diff}
\setlength{\belowrulesep}{-.5pt}
\setlength{\aboverulesep}{-.3pt}
\begin{center}
\begin{tabular}{r rrr rrr rrr }
  \hline
Model &\multicolumn{3}{c}{all cells} 
&\multicolumn{3}{c}{cells s. t. $F_k>3$}  & \multicolumn{3}{c}{cells s. t. $F_k\leq 3$}\\
\cmidrule(lr){2-4}\cmidrule(lr){5-7} \cmidrule(lr){8-10}
 & sign. & abs. & sq. & sign. & abs. & sq. & sign. & abs. & sq. \\ 
 \hline
NP+I & -7.9 & 80.5 & 38.7 & 31.4 & 38.7 & 13.3 & -39.3 & 41.8 & 25.4 \\ 
NP+II& -15.3 & 81.4 & 40.2 & 24.8 & 37.6 & 12.6 & -40.1 & 43.8 & 27.6 \\ 
P+I & -17.4 & 94.7 & 51.6 & 27.8 & 46.3 & 19.2 & -45.2 & 48.4 & 32.4 \\ 
P+II& -9.6 & 84.7 & 41.8 & 29.6 & 41.6 & 14.9 & -39.2 & 43.1 & 26.9 \\ 
\hline
  &\multicolumn{3}{c}{} 
&\multicolumn{3}{c}{cells s. t. $F_k>10$}  & \multicolumn{3}{c}{cells s. t. $F_k\leq 10$}\\
\cmidrule(lr){5-7} \cmidrule(lr){8-10}
 &  & & & sign. & abs. & sq. & sign. & abs. & sq. \\ 
 \cmidrule(lr){5-10}
   NP+I       &   &  & & 18.1 & 18.1 & 6.8 & -26.0 & 62.4 & 32.0 \\ 
   NP+II      &  &  &  & 12.9 & 13.9 & 4.6 & -28.1 & 67.5 & 35.6 \\ 
  P+I             &  &  &  & 18.1 & 19.8 & 9.8 & -35.5 & 75.0 & 41.8 \\ 
   P+II          &  & &  & 14.3 & 15.3 & 5.2 & -23.9 & 69.4 & 36.6 \\ 
\hline
\end{tabular}
\end{center}
\vspace{\SPACE}
\end{table}
Analyses of the performance of estimators of $\tau_{2,k}^*$s,  reported in Table~\ref{tab:diff}, show that, going from the NP+I to the NP+II model, the improvement of per-cell risk estimates for low risk cells (large $F_k$s) tends to be greater than the improvement of per-cell risk estimates for high risk cells  (small $F_k$s), a fact that may explain the increasing negative bias observed at global level in  Table~\ref{tab:results3}. 
  An analogous suggestion comes from Figure~\ref{boxplot:tau1k}, concerning estimators of  
 ${\tau}_{1,k}^*$s.    Figure~\ref{boxplot:tau1k} presents the boxplots of estimates of  ${\tau}_{1,k}^*$  restricted to cells which are population uniques ($F_k=1)$. It shows that the two-way interactions models, NP+II and P+II, are by far superior to the nonparametric independence model, NP+I, on those cells. Therefore, from this Figure we can conclude that  the 
 worse performance at global level of  the NP+II model  observed in  Table~\ref{tab:results3},  not only  
 in comparison with the NP+I model but also with the P+II model, 
is an unpleasant consequence of the the greater improvement achieved by the nonparametric all-two-way interaction model  NP+II 
on cells where the true risk $\tau_{1k}$ is zero ($F_k>1)$. Further evidence about this fact is given in Figure~\ref{fig:tau2k:IIvsII}.
%
  In other words, these analyses reveal a trade-off between good global risk estimates and good  {\it local} (per-cell) risk estimates, under a log-linear modelling of the Poisson parameters, which sounds like a negative result in the field of disclosure risk estimation. Indeed, under the nonparametric   independence model, this is the natural consequence of a less detailed modelling of the cell parameters $\lambdavect$ that, in sparse tables, would otherwise be unidentifiable and, therefore, unestimable. 
 From such trade-off, however, we can draw valuable indications about modelling in different fields. For instance, in  cases where  the focus is on accurate  estimation of total counts or rates (more sensitive to accurate estimates of large $F_k$s) based on sample unique cells, from 
the analyses reported in Table~\ref{tab:diff} (see also  Figure~\ref{fig:tau2k:IIvsII}, bottom row)
 we can expect  that the NP+II model may outperform the P+II model, returning smaller estimation errors.  
 In this respect, the above mentioned decrease of the s.e. is an interesting result and deserves further analyses (a first analysis is provided in the supplementary material A.4). We are currently working to show that all these features extend to sample cells including a few (or no) observations, in order to exploit this result in small area  estimation problems, where typical contingency tables are not so large and sparse as in  disclosure risk estimation. \\

In conclusion, through a long journey around different forms and sources of bias, we gained  a new perspective on model selection for disclosure risk estimation yielding an effective,  simple method also able to face the challenging issue, rarely treated in the literature, of the selection-induced  bias. In addition, we acquired new solid suggestions about modelling in small area estimation problems.

\newpage

\bibliographystyle{ba}
\bibliography{Bibliography-MM-MC,biblio,filippone}

\newpage

\bigskip
\begin{center}
{\large\bf SUPPLEMENTARY MATERIAL}
\end{center}

\section*{A.1. Illustrative tables and models used in  Sections~\ref{sec:npcorrection} and~\ref{sec:modsel}}
In this material we provide details about the data and models that were used in the illustrative examples summarized in  Sections~\ref{sec:npcorrection} and~\ref{sec:modsel} of the article.   

   First, we reconsider the same table of 3,600,000 cells used in \citet{mvreiter:jasa12}.  A description of the ten key variables generating such ``large'' California table is provided in Table~\ref{tab:CALtabs} (left panel).  It is built  from the 5\% public use microdata sample of the U.S. 2000 census for the state of California \citep[IPUMS][]{IPUMS}; the set of $N=1,150,934$ individuals aged 21 and over is taken as the reference population. In addition,  two new contingency tables are obtained by global suppression of variables DISAB and  VETST, yielding a ``medium'' table of 900,000 cells,  and  global suppression of variable INCR,  yielding a ``small'' table of 360,000 cells. Clearly, global suppression is not applied here as a protection strategy, but rather as a way to create contingency tables with  different characteristics. 
 
Since by design the previous  tables do not include structural zeroes,  we also consider a  contingency table of 844,800 cells, half of which are structurally empty. The data come from the set of $N=794,986 $ individuals recorded in the 7\% public use microdata sample of the Italian National Social Security Administration, 2004 (source: Work Histories Italian Panel, WHIP), treated here as the  population; these records are cross-classified according to the eight key variables  listed in  Table~\ref{tab:CALtabs} (right panel). 
 In all cases we draw random samples with fraction $\pi=0.05$. 

 \begin{table}[th]
\caption{Key variables under consideration (number of categories in parentheses) and their labels in the California  (left) and  WHIP data (right).}
\label{tab:CALtabs}
\begin{center}
{ 
\begin{tabular}{llll} 
\hline
\multicolumn{2}{c}{ \em Large California } &\multicolumn{2}{c} {\em WHIP}   \\
\cmidrule(lr){1-2}  \cmidrule(lr){3-4}
\multicolumn{1}{c}{Label} &\multicolumn{1}{c}{Variable} &  \multicolumn{1}{c}{Label} &\multicolumn{1}{c}{Variable} \\ 
\cmidrule(lr){1-1} \cmidrule(lr){2-2}\cmidrule(lr){3-3} \cmidrule(lr){4-4}
CHIL & number of children (10)           & AORIG & area of origin (11)   \\ 
AGE & age (10)          & AGE & age (12)    \\
SEX & sex (2)          & SEX & sex (2)   \\ 
MARST & marital status  (6)     & RWORK  & region of work (20)  \\ 
RACE & race  (5)     & ESEC & economic sector (4)   \\ 
EDU & education   (5)   & WAGF & wage guaranteed fund (2)   \\ 
EMPST & employment status  (3)    & WORKP & working position (4)   \\ 
INCR & income (10)   & FSIZE &  firm size (5)  \\ 
DISAB & disability (2)       & &   \\ 
VETST & veteran status (2)       &  &   \\ 
\hline
\end{tabular}
}
\vspace{\SPACE}
\end{center}
\end{table}
For each of the four contingency tables just described, we consider the  models  listed  in Table~\ref{tab:ModSETS}. The nonparametric (NP) models are selected  as illustrated at the beginning of  Section~\ref{sec:npcorrection} and formally described in Section~\ref{sec:modsel}. 
%
Often, for comparison, we also consider the parametric counterparts of such models, labelled P. 
The latter are Bayesian log-linear models with the same fixed effects
 \newpage
  \begin{table}[hb]
\caption{Log-linear models for California and WHIP tables: model label, structure of fixed effects (parametric component), prior on the random effects,  and number  of additional parameters compared to those included in the independence model.}
\label{tab:ModSETS}

\begin{tabular}{p{1.50cm}p{5cm}p{1.8cm}p{1.8cm}}
\hline
Label &	Shorthand for fixed effects &  \splitcell{Prior for\\random effects}  & \splitcell{n. of extra\\  parameters}   \\ 
\hline
$California\,\, Large$ &  &   \\ 
\hline
NP$_a$ & I + SEX*VETST            & ${\cal{D}}(m, Ga(a,b))$  & 1\\ 
NP$_b$ & I + EMPST*INCR           & ${\cal{D}}(m, Ga(a,b))$   & 18 \\ 
NP$_c$ & I+ SEX*VETST + EMPST*INCR          & ${\cal{D}}(m, Ga(a,b))$  & 19 \\
P$_a$ & I + SEX*VETST            &  $Ga(a,b)$ & 1\\ 
P$_b$ & I + EMPST*INCR           & $Ga(a,b)$   & 18 \\ 
P$_c$ & I+ SEX*VETST + EMPST*INCR          & $Ga(a,b)$   & 19 \\
\hline
$California\,\, Medium$ &  &   \\ 
\hline
NP$_a$ & I + EMPST*SEX          & ${\cal{D}}(m, Ga(a,b))$  & 3 \\ 
NP$_b$ & I + EMPST*INCR        & ${\cal{D}}(m, Ga(a,b))$  & 18\\
NP$_c$ & I + EMPST*SEX  + EMPST*INCR  & ${\cal{D}}(m, Ga(a,b))$  & 21 \\ 
NP+I & I&$ {\cal{D}}(m, Ga(a,b))$  & -\\
P$_a$ & I + EMPST*SEX          & $ Ga(a,b)$  & 3 \\ 
P$_b$ & I + EMPST*INCR        & $Ga(a,b)$  & 18\\ 
 P$_c$ & I + EMPST*SEX  + EMPST*INCR  & $Ga(a,b)$  & 21 \\
P+I & I& $Ga(a,b)$  & -\\
 \hline
$California\,\, Small$ &  &   \\ 
\hline
NP$_a$ & I + SEX*VETST        &${\cal{D}}(m, Ga(a,b))$  & 1\\ 
NP$_b$ & I + MARST*RACE  & ${\cal{D}}(m, Ga(a,b))$  & 20\\  
NP+I & I&$ {\cal{D}}(m, Ga(a,b))$  & -\\
P$_a$ & I + SEX*VETST        & $Ga(a,b))$ & 1\\
P$_b$ & I + MARST*RACE  & $Ga(a,b))$  & 20\\  
P+I & I& $Ga(a,b)$  & -\\
\hline
$WHIP$ &  &   \\ 
\hline
NP$_a$  & I + ESEC*WORKP        & ${\cal{D}}(m, Ga(a,b))$ & 9\\ 
NP$_b$ & I + ESEC*FSIZE        & ${\cal{D}}(m, Ga(a,b))$ & 12\\ 
NP$_c$  & I + ESEC*WORKP + ESEC*FSIZE       & ${\cal{D}}(m, Ga(a,b))$ & 21\\ 
NP$_d$  & I + ESEC*WORKP + ESEC*SEX + ESEC*WAGF + ESEC*FSIZE        & ${\cal{D}}(m, Ga(a,b))$  & 27 \\ 
NP$_e$  & I + ESEC*WORKP + ESEC*SEX + ESEC*WAGF + AGE*WORKP   &  ${\cal{D}}(m, Ga(a,b))$& 48\\ 
NP+I & I&$ {\cal{D}}(m, Ga(a,b))$  & -\\
P+I & I& $Ga(a,b)$  & -\\
\hline
\end{tabular}
\vspace{\SPACE}
\end{table}
 and parametric (Gamma distributed, as it is explained later) random effects: they represent the special case of NP models  for  an indefinitely large $m$. 
As recalled in Section~\ref{intro1}, in practice such parametric counterparts are equivalent to  log-linear models without random effects (\ref{formula2}), since the corresponding 
 risk estimates are nearly identical  \citep[]{elamir:skinner}.

Using the fact that, conditionally on the random effects ($NP$ or $P$), our models differ 
only for the specification of the vector  $\betavect$, we further distinguish them by referring to their parametric component. 
For instance, as  we use the shorthand notation NP+I  to denote the nonparametric model whose fixed effects include the main effects of all key variables (nonparametric independence model); likewise, 
we denote  by NP+I+A*B  the nonparametric model whose fixed effects additionally include  the two-way interaction parameters between all levels of key variables A and B. 
Therefore, a comprehensive  description of all models explored in  Sections~\ref{sec:npcorrection} and~\ref{sec:modsel} 
can be read in  columns of  Table~\ref{tab:ModSETS}: model label (dictated by the  nature of random effects), shorthand for fixed effects included in the model, prior on random effects and number, $d$, of extra parameters, implied by the interaction terms (e.g. A*B) added to  the independence model.

In all of these applications  we reparametrize the random effects so that $\omega_k=e^{\phi_k}$ is drawn from a Dirichlet process with  base measure $Ga(a, b)$  (where $b$ is the rate parameter), thereby extending the model defined in \citet{elamir:skinner},  and assume a Gamma-distributed precision $m$. The hyperparameters are fixed so as to specify vague priors: for the random effects we take $a=1$, $b=0.1$;  
as regards the fixed effects $\betavect$, we consider a vague Gaussian prior, $N(\boldmath{0},I\sigma^2)$; finally, we take $m\sim \mathrm{Gamma}(1,0.1)$. 

\section*{ A.2. Analysis of per cell risk estimates under nonparametric vs. parametric independence  models }

\begin{figure}[t]
  \centering
\def\arraystretch{-1}
\begin{tabular}{ccc}
 (a)&(b)&(c)\\
 \includegraphics[height=0.3\textwidth, angle=-90]{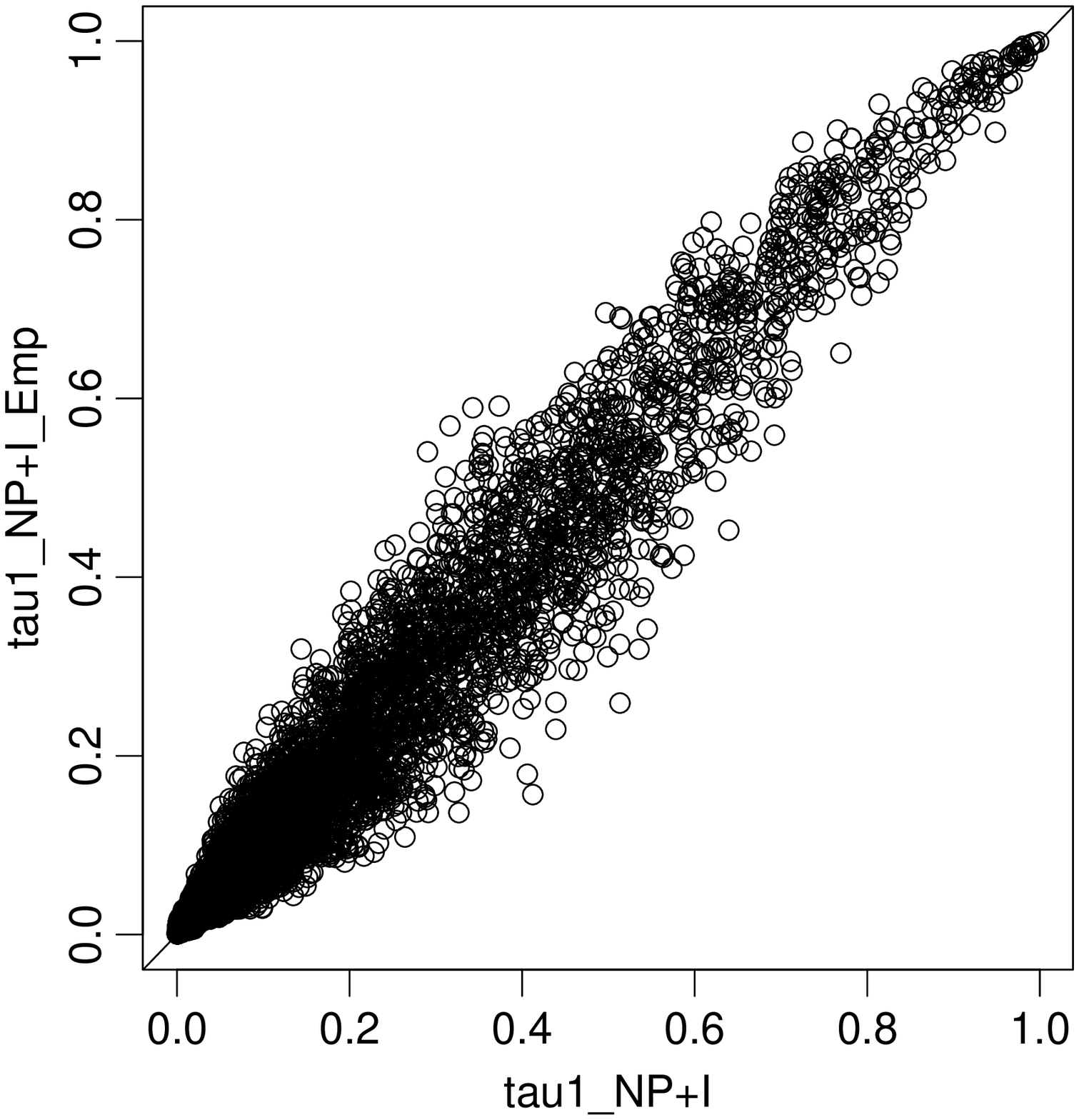} &
  \includegraphics[height=0.3\textwidth,angle=-90]{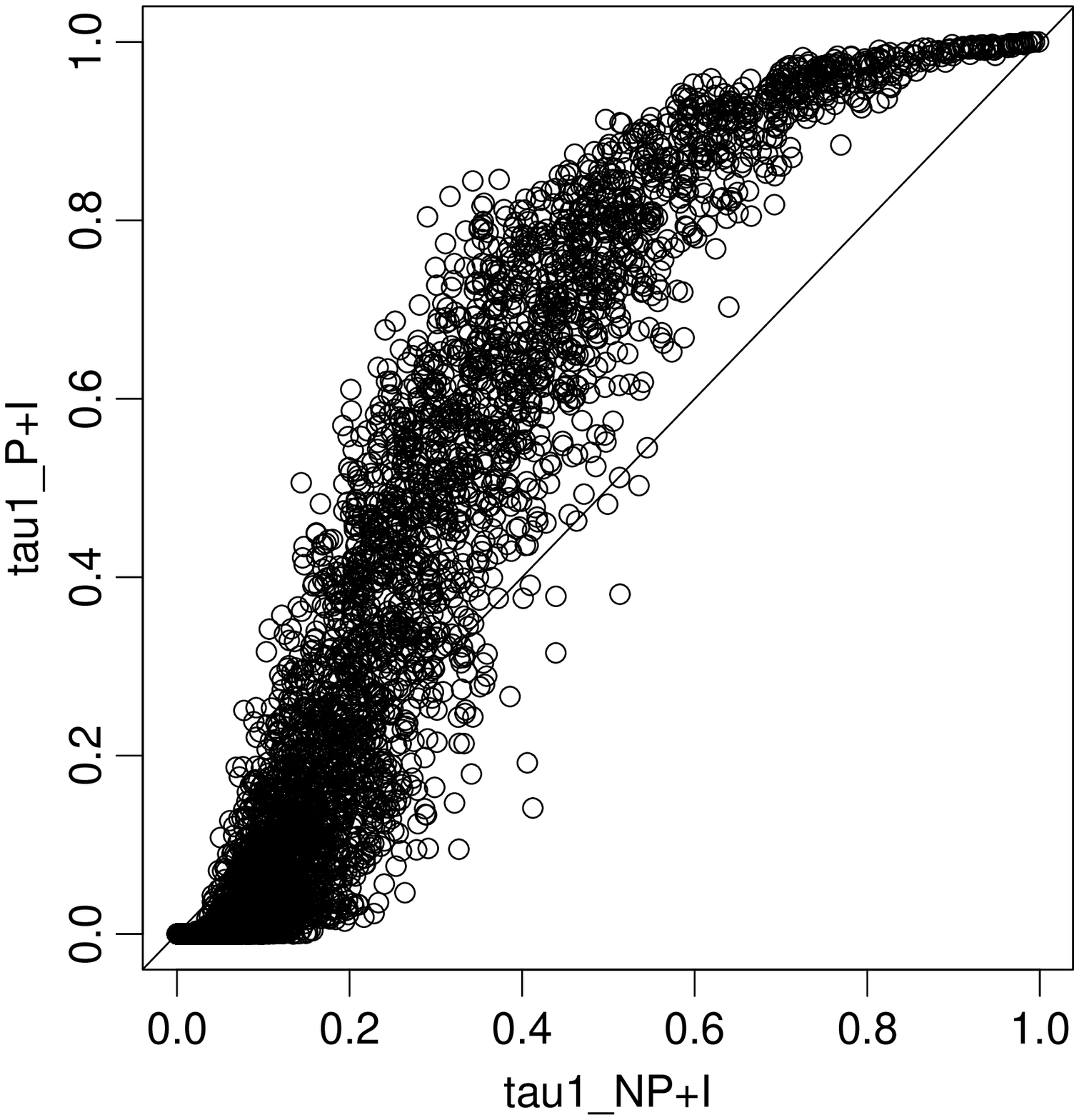} &
  \includegraphics[height=0.3\textwidth,angle=-90]{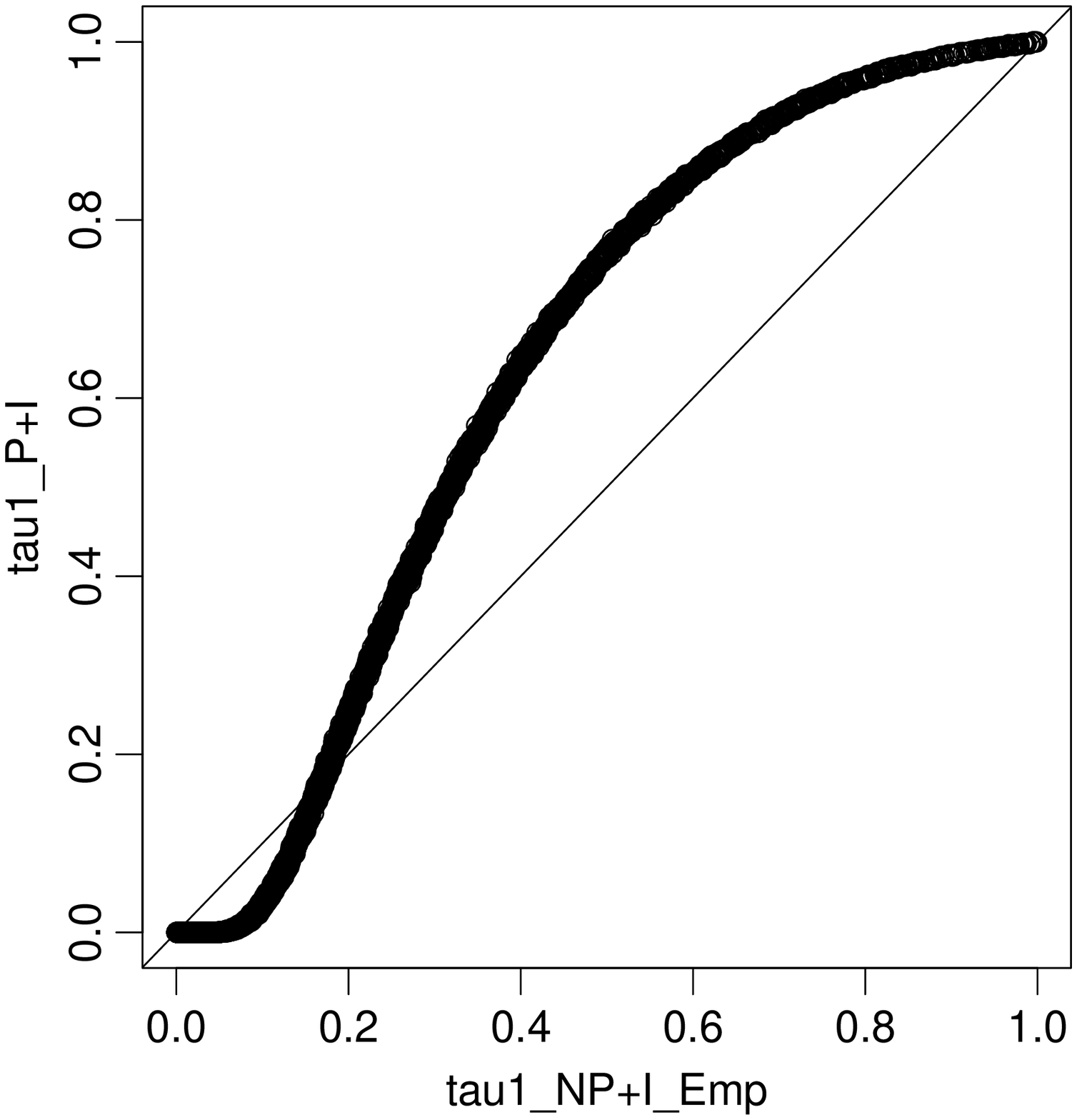}  \\
 \includegraphics[height=0.3\textwidth, angle=-90]{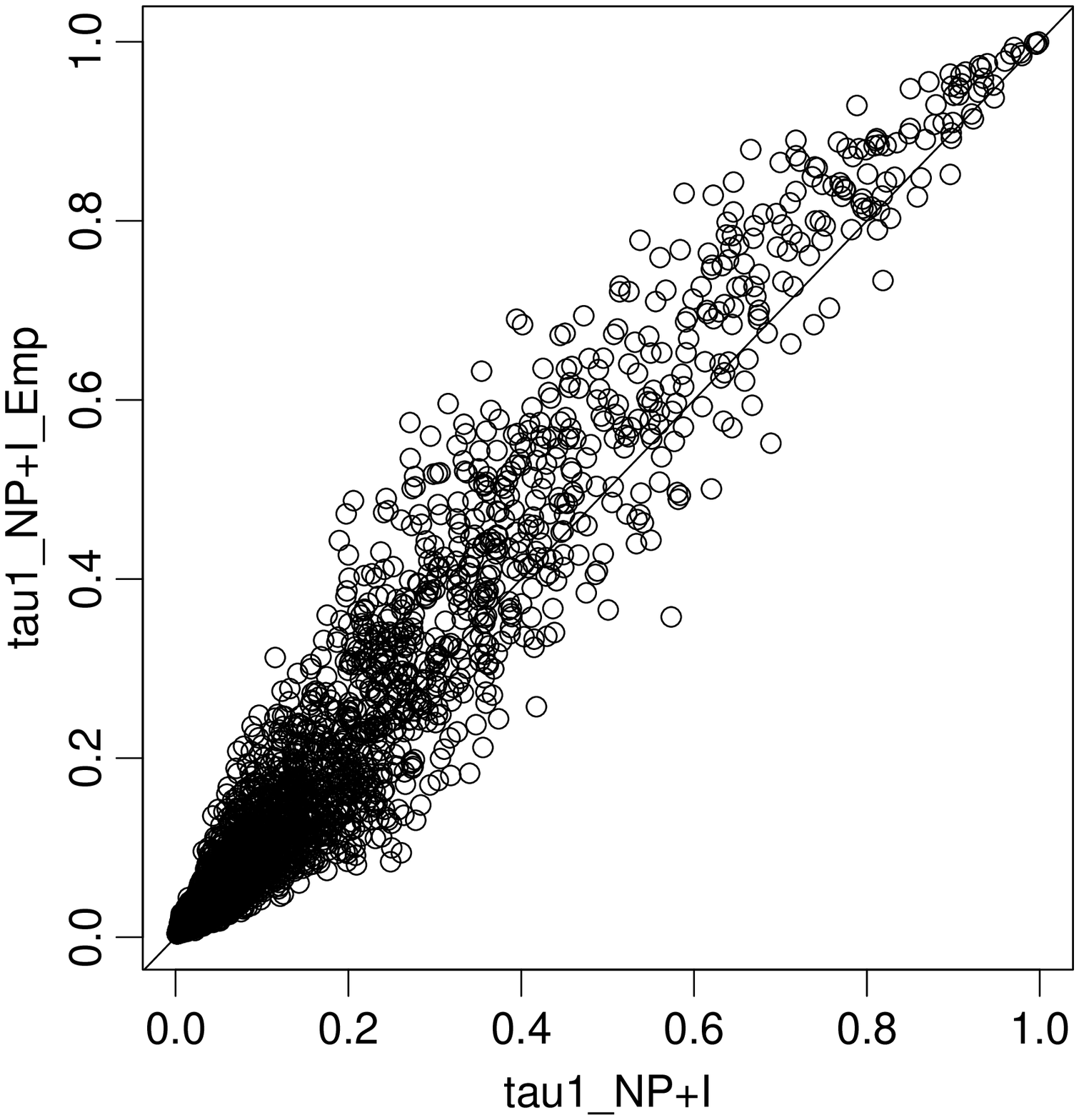} &
  \includegraphics[height=0.3\textwidth,angle=-90]{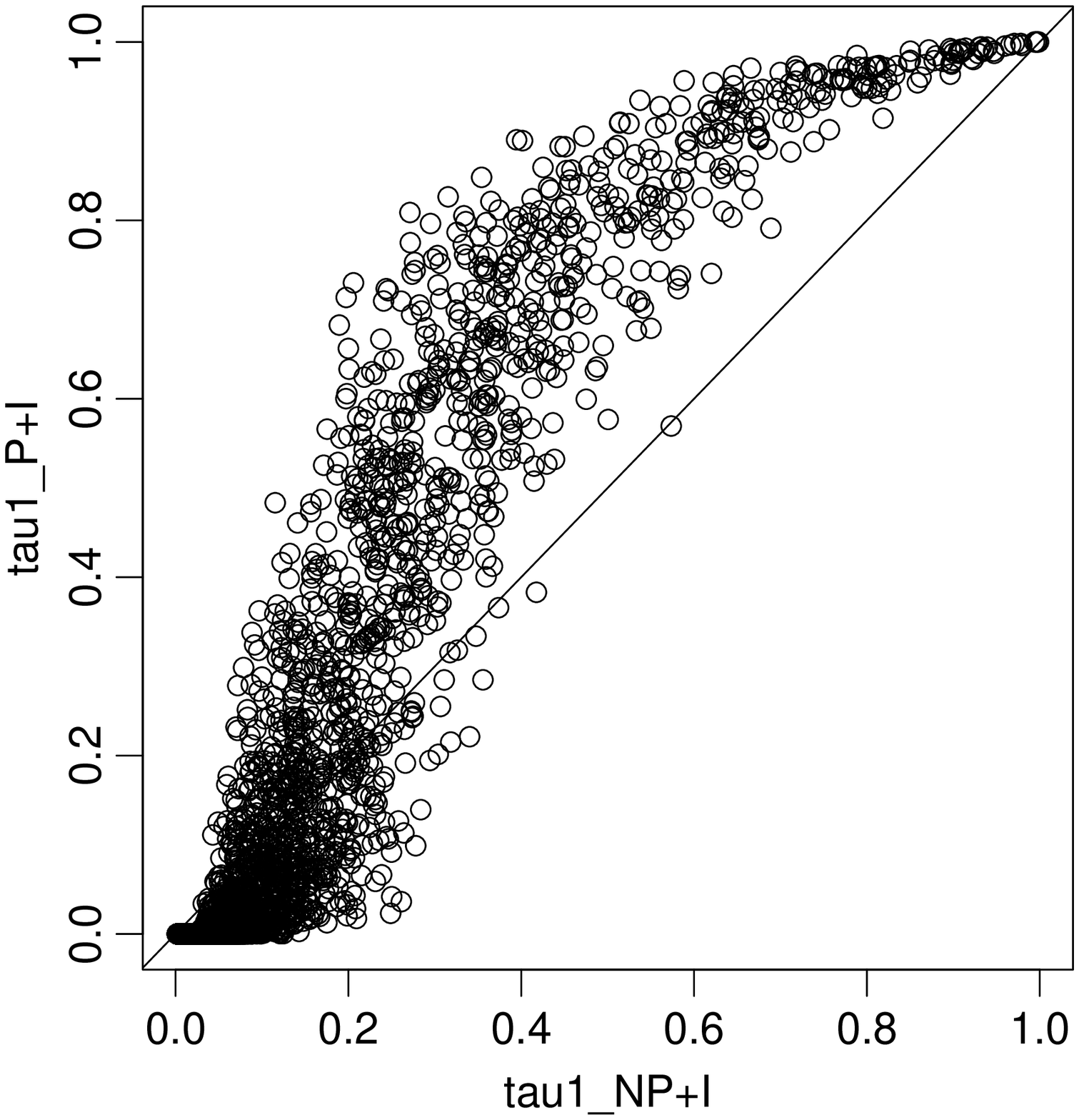} &
  \includegraphics[height=0.3\textwidth,angle=-90]{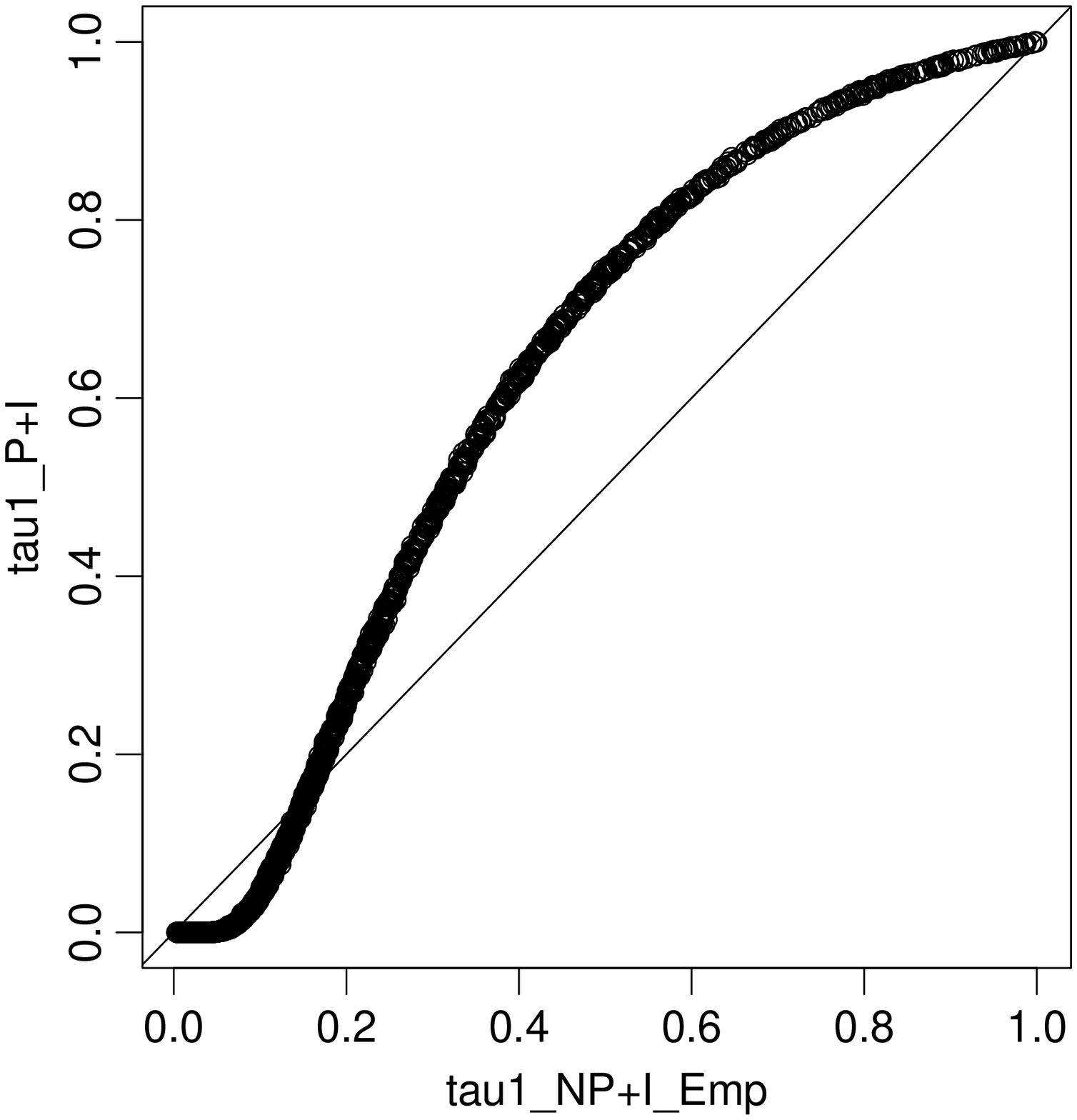} \\
    \includegraphics[height=0.3\textwidth, angle=-90]{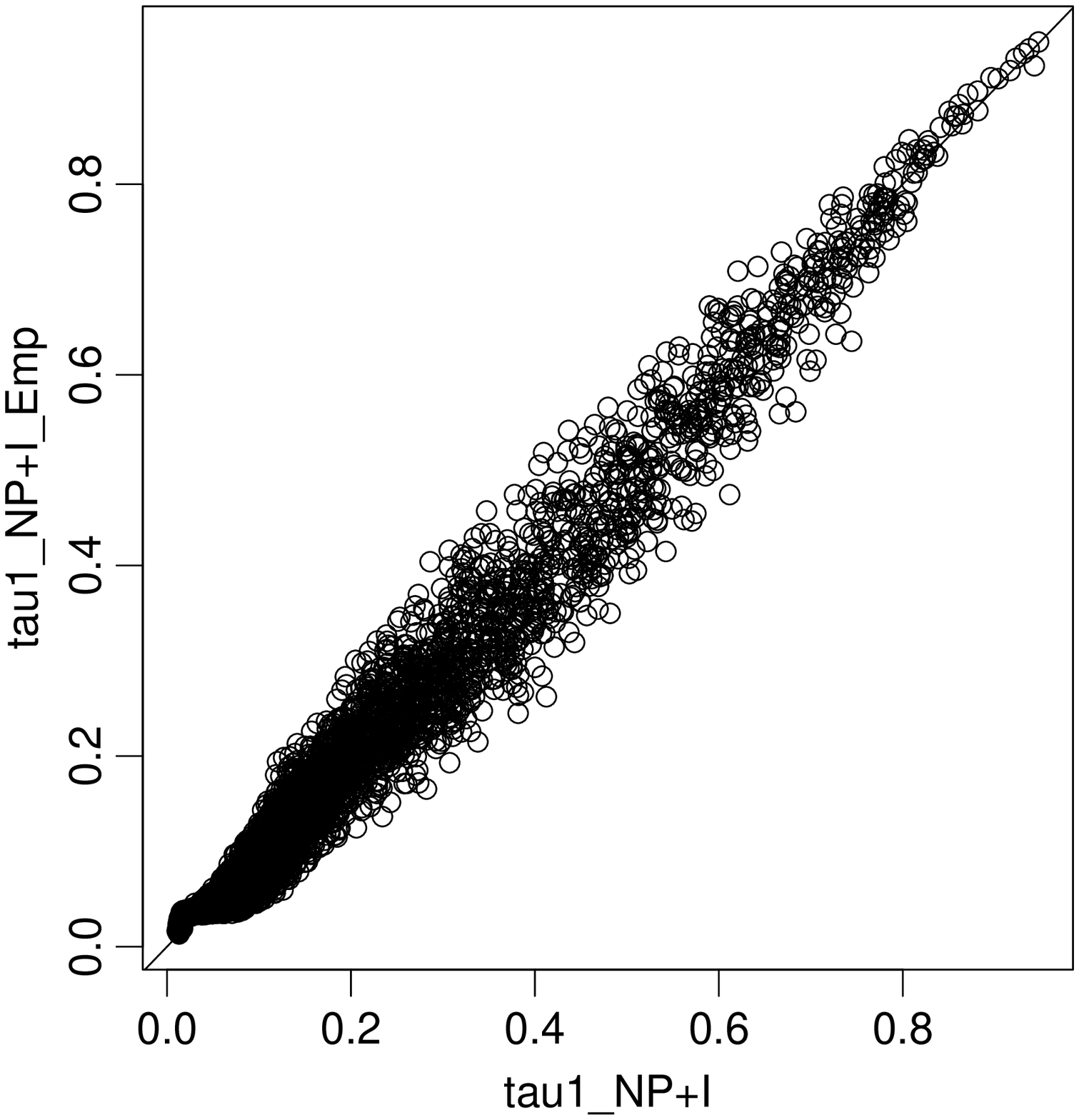}& 
      \includegraphics[height=0.3\textwidth,angle=-90]{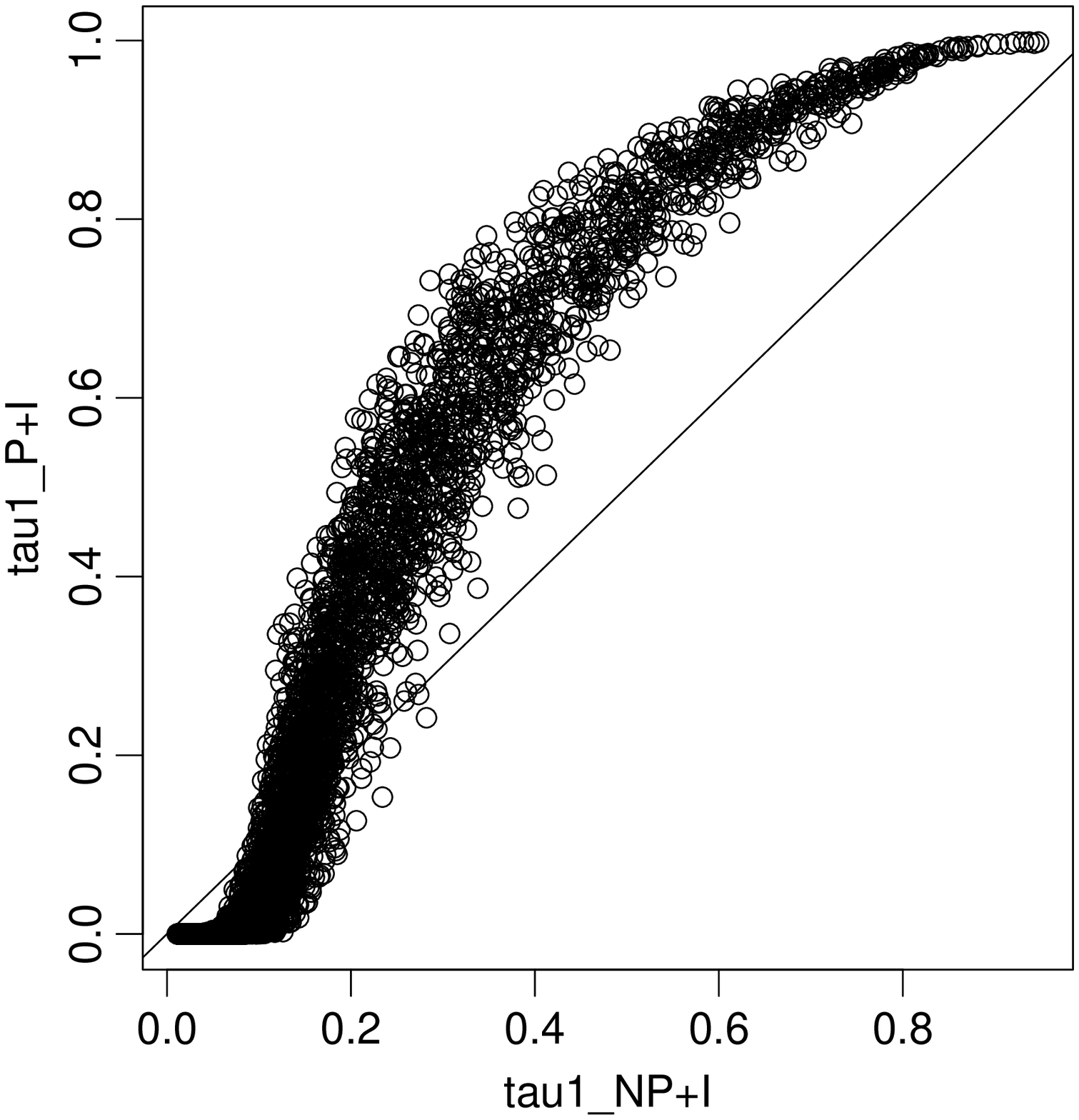} &
      \includegraphics[height=0.3\textwidth,angle=-90]{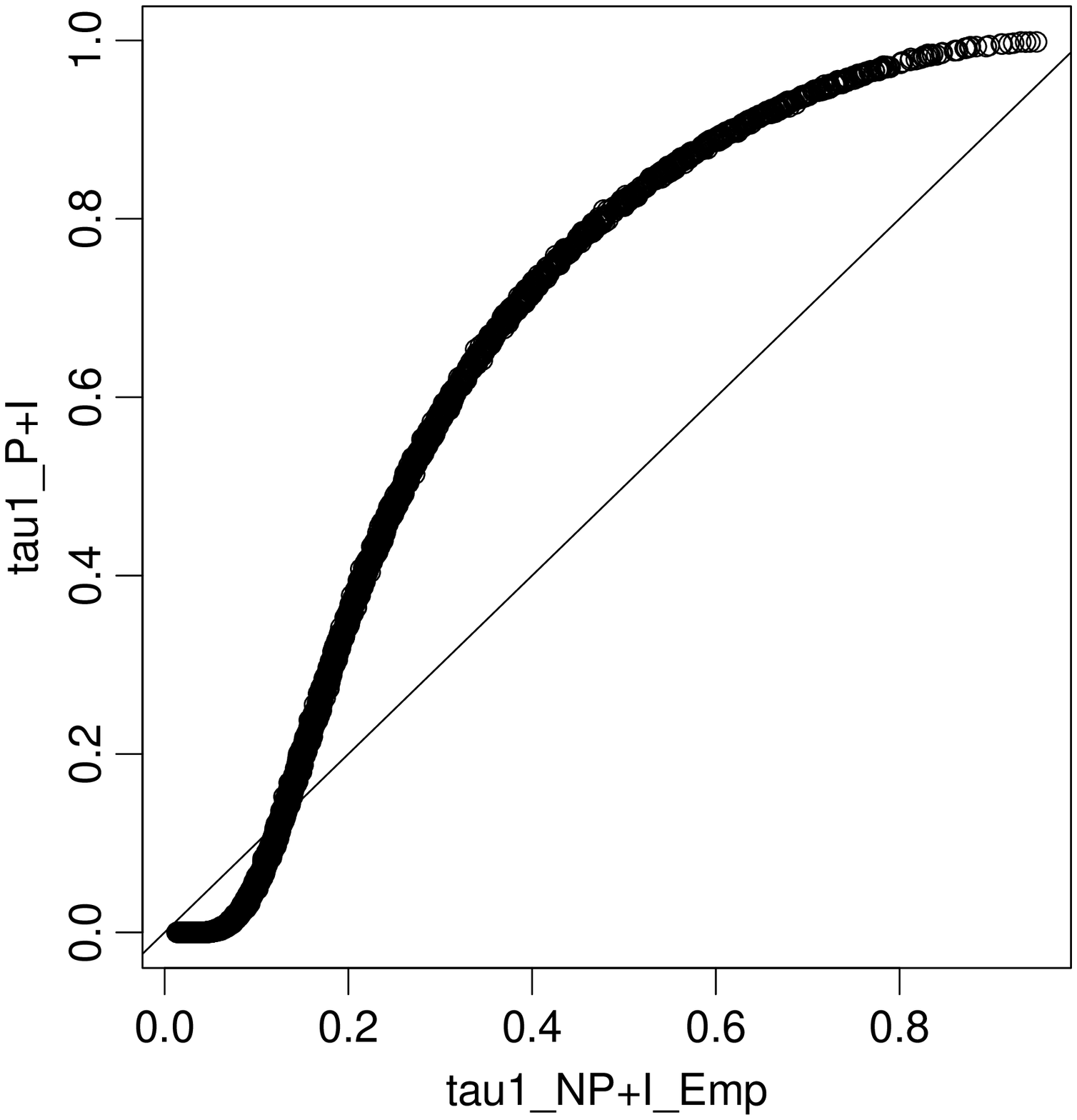} 
\end{tabular}
 \caption{ {\it California Medium, California Small and WHIP} tables (first, second and third row, respectively). For each of the three tables, the figures report estimates of per cell risks,  $\hat{\tau}_{1,k}$,  under the nonparametric independence model (horizontal axis) and under its parametric counterpart, in the presence of a vague Gaussian prior on ${\betavect}$ (column b), or a degenerate prior  putting mass at the ML estimate of ${\betavect}$ (column c). The latter estimates are referred to as nonparametric empirical Bayesian estimates, and the corresponding model is denoted by  the additional subscript ``Emp". For the sake of completeness, column (a) reports both the nonparametric and nonparametric empirical Bayesian per cell risk estimates of  ${\tau}_{1,k}$.  } 
 \label{fig:sigmoidinuovi:tab1}
 \end{figure}

Figure~\ref{fig:sigmoidinuovi:tab1}  presented in this material compares the estimates of $\tau_{1,k}$ obtained under the  nonparametric independence model NP+I and under its  parametric counterpart  P+I, in three different tables: California Medium, California  Small and Whip. We  can observe different sigmoidal shapes: two of them (first two rows) describe corrections of parametric per-cell risk estimates sufficient to obtain  good nonparametric  estimates of global risks (see Figure~\ref{fig:quantili:tab1e2}); the sigmoidal shape in the third row,  though much more pronounced, does not achieve the same result (see Figure~\ref{fig:whipmidtau1quantiles}).

\section*{A.3. Illustrative table and models used in Section~\ref{sec:tab:piccole1} }

In Section~\ref{sec:tab:piccole1}  we perform a detailed analysis of per cell risk estimates, comparing the estimation bias associated to more and less parsimonious nonparametric models. We refer to a  small and dense contingency table,  called the Small WHIP table, that, once again, is obtained from the 7\% microdata sample of the Italian National Social Security Administration, 2004. 
%
This time we treat  the  $N=450,238$ individuals whose workplace falls into 4 specific geographic areas as the reference population, from which we draw a random sample with fraction {$\pi = 0.1$}. 
The table has the following spanning (key) variables 
(number of levels in parentheses): geographic area~(4), sex~(2), age~(11), ethnicity~(5), and economic activity~(9), returning a total of $K=3,960$ cells.  

We consider as ``reference'' models (representative of more and less parsimonious models)  the nonparametric independence and the all-two-way interactions models,  NP+I and NP+II, respectively, also estimating  their parametric counterparts, P+I and P+II, for comparison.
In this application the base measure of the DP prior for random effects is  $G_0=\Norm(\alpha,\sigma^2)$, 
which extends  the model in \citet{skinner:holmes}; we assume $\alpha \sim \Norm(0, 10)$, $\sigma^2\sim \mathrm{invGamma}(1,1)$ and $m \sim  \mathrm{Gamma}(1,1)$. Finally, we take a reasonably vague Gaussian prior $\Norm(0, 10 I)$ on $\betavect$.

\section*{A.4. Some remarks on standard errors of global risk estimators}
Here  we try to explain why the standard error decreases when going from the  NP+I model to the  NP+II model, and viceversa increases when going from the P+I to the P+II modeI. If we consider the deviances of the per-cell risk estimates $\hat{\tau}^*_{i,k}$ around their mean $\hat{\tau}^*_{i}/K$ (hereafter denoted by $D_b$) for the parametric and nonparametric all-two-way interactions models, we obtain very similar values and, given that the $\hat{\tau}^*_{i,k}$ are in turn means within each cell, from 
\begin{equation} 
(s.e.(\hat \tau_i^*))^2=\frac{1}{H}\sum_h^H\Big(\sum_k^K \tau_{i,k}^{*(h)}-\hat{\tau}_{i}^{*}\Big)^2=V_w+D_b+C_b,
\end{equation}
where 
\begin{equation}
V_w=\sum_k^K \frac{1}{H}\sum_h^H (\tau_{i,k}^{*(h)})^2- \sum_k^K (\hat{\tau}_{i,k}^{*})^2
\end{equation}
is the sum of the variances within each cell,
\begin{equation}
D_b=\sum_k^K (\hat{\tau}_{i,k}^{*})^2 - K\Big(\frac{\hat{\tau}_{i}^{*}}{K}\Big)^2
\end{equation}
is the deviance between cells and 
\begin{equation}
C_b=\sum_k^K \sum_{j \neq k} ^K \frac{1}{H}\sum_h^H \tau_{i,k}^{*(h)} \tau_{i,j}^{*(h)}- K (K-1)\Big(\frac{\hat{\tau}_{i}^{*}}{K}\Big)^2
\end{equation}
is the sum of codeviances between cells, 
we can conclude that the smaller standard error observed in Table~\ref{tab:results3} under the  NP+II model compared to the one under the P+II model is due to smaller values of the variances within cells and/or of codeviances between per-cell risks.
If, in addition, we consider a nonparametric model without fixed effects -under which $\hat{\tau}^*_{1}=16.5 (4.5)$ and  $\hat{\tau}^*_{2}=76.4 (7.4)$-, where the  components $V_w$ and $C_b$ are essentially the only relevant components  of the s.e. being $D_b<0.001$, we can affirm that, as the complexity of the nonparametric  log-linear model increases,  the decrease of $V_w$  and/or $C_b$ prevails over the slight increase of $D_b$. Vice versa, going from the parametric independence model P+I to the all two-way interactions model P+II, the component $D_b$ slightly decreases and is overwhelmed by the increase of  $V_w$  and/or $C_b$. In this respect, consider that under
 parametric models the only way to increase the association between cells is by means of the introduction of further interaction terms.

\section*{B. Implementation of the MCMC approach}

The Markov Chain Monte Carlo (MCMC) sampler employed here is a Gibbs sampler, where groups of parameters are sampled one after the other. 
In particular, the sequence of MCMC steps amounts in drawing samples from the conditionals $\betavect | \mathrm{rest}$, $\phivect | \mathrm{rest}$, and $m | \mathrm{rest}$.
Samples from the posterior distribution over $\betavect$, $\phivect$, and $m$ allows one to estimate per-cell risks through Monte Carlo averaging.

{\bf Sampling $\betavect$} --
The conditional distribution of $\betavect | \mathrm{rest}$ is not of known form, given that the prior on $\betavect$ is Gaussian and the likelihood is Poisson. 
Therefore, we employ Metropolis-within-Gibbs samplers, where a proposal is accepted or rejected according to a Metropolis ratio \citep{Roberts09}; these can include, e.g., Metropolis-Hastings \cite{Metropolis53} or Hybrid Monte Carlo \cite{Duane87,Neal93}, but in this work we employ the so-called Simplified Manifold Adjusted Langevin Algorithm (SMMALA) \citep{Girolami11}. 
SMMALA is one instance of manifold MCMC methods, which are characterized by the fact that they exploit the curvature of the log-likelihood, allowing for efficient moves in the parameter space. 
SMMALA has been shown to be effective for problems similar to the ones considered here, where the posterior is unimodal and is not characterized by strong skewness.  
SMMALA approximates the diffusion on the statistical manifold characterizing $p(f_1, \ldots, f_K | \betavect, \mathrm{rest})$.
Defining $M$ to be the metric tensor obtained as the Fisher Information of the model plus the negative Hessian of the prior, and $\epsilon$ to be a discretization parameter, SMMALA can be thought of as a Metropolis-Hastings sampler with a position-dependent proposal.
The curvature of the log-likelihood determines the step-size of the proposal through the metric tensor $M$ as follows $p(\betavect^{\prime} | \betavect) = N(\betavect^{\prime} | \muvect, \epsilon^2 M^{-1})$, with $\muvect = \betavect + \frac{\epsilon^2}{2}  M^{-1} \nabla_{\betavect} \log[p(f_1, \ldots, f_K | \betavect, \mathrm{rest})]$.
The complexity of the update is $\mathcal{O}(K D^3)$ where $K$ is the number of cells and $D$ is the size of $\betavect$; the linearity in $K$ makes it well suited in applications where the number of cells is large, while the cubic scaling in $D$ makes it suitable for models with a small number of $\betavect$ parameters.

{\bf Sampling $\phivect$} --
In  Sections~\ref{sec:npcorrection} and~\ref{sec:modsel} of this work, we exploit the conjugacy between the base Gamma measure and the Poisson likelihood to derive an efficient sampler for $\phivect$.
We implemented the MCMC sampler ``Algorithm~3'' proposed in the review paper of MCMC methods for DP models in \citet{Neal00}. 
In a nutshell, we choose a 
distribution for $G_0$ 
such  that $\omega=e^{\phi}$ is given a Gamma base measure, for which we can exploit conjugacy with the Poisson likelihood. 
A similar argument holds when $\phi$ is given the $IG$ distribution. 
This allows us to integrate out the values of $\phi$ analytically $ \int p(f_k|\mathbf \betavect, \phi) dG_0(\phi)$, where we expressed $p(f_k|\mathbf \betavect, \phi)$ as the likelihood for a single point.
As a result, it is possible to derive a sampler that allocates cells to an unknown number of clusters and to draw directly a value for the random effect for each cluster.
The complexity of the update is $\mathcal{O}(K)$.\\
In Section~\ref{sec:tab:piccole1}, where the base measure is not conjugate with the Poisson likelihood, we adopt the ``Algorithm~5''  in \citet{Neal00}, as it easy to implement and as it achieves satisfactory performance in the given application. 

{\bf Sampling $m$} --
We choose a Gamma prior for the $m$ parameter. 
With this choice, it is possible to draw samples from the posterior distribution over $m | \mathrm{rest}$ directly following \citet{Escobar:West}.






\end{document}